\documentstyle[aps,eqsecnum,prb,preprint,tighten,amsfonts,epsf]{revtex}
\begin{document}
\preprint{IC/97/82}
%
%
%
%
%

\title{LOW--TEMPERATURE REGIMES AND FINITE SIZE--SCALING IN
A QUANTUM SPHERICAL MODEL}
\date{July 6, 1997}
\author{Hassan Chamati$^1$\thanks{
Permanent Address:
Institute of Solid State Physics, Bulgarian Academy of
Sciences, Tzarigradsko chauss\'ee 72, 1784 Sofia, Bulgaria.}
Daniel M. Danchev$^2$, Ekaterina S. Pisanova$^3$
and Nicholay S. Tonchev$^4$}
\address{$^1$International Centre for Theoretical Physics, Trieste, Italy}
\address{$^2$Institute of Mechanics, Bulgarian Academy of Sciences,\\
Acad. G. Bonchev, bl. 4, 1113 Sofia, Bulgaria.}
\address{$^3$University of Plovdiv, Faculty of Physics, 24 ``Tzar Assen''
str.,\\
4000 Plovdiv, Bulgaria.}
\address{$^4$Institute of Solid State Physics, Bulgarian Academy of
Sciences, \\
Tzarigradsko chauss\'ee 72, 1784 Sofia, Bulgaria.}

\maketitle

\begin{abstract}
A $d$--dimensional quantum model in the spherical approximation confined
to a general geometry of the form $L^{d-d^{\prime}}\!\!\times\!\!\infty^{d^{
\prime}}\!\!\times\!\! L_{\tau}^{z}$ ($L$--linear space size and
$L_{\tau}$--temporal size) and subjected to periodic boundary conditions
is considered. Because of its close relation with the quantum rotors model it
can be regarded as an effective model for studying the low--temperature behavior
of the quantum Heisenberg antiferromagnets. Due to the
remarkable opportunity it offers for rigorous study of finite--size effects
at arbitrary dimensionality this model may play the same role in quantum
critical phenomena as the popular Berlin--Kac spherical model in classical
critical phenomena. Close to the zero--temperature quantum critical point,
the ideas of finite--size scaling are utilized to the fullest extent for
studying the critical behavior of the model. For different dimensions $1<d<3$
and $0\leq d^{\prime}\leq d$ a detailed analysis, in terms of the special
functions of classical mathematics, for the free energy, the susceptibility
and the equation of state is given. Particular attention is paid to the
two--dimensional case.
\end{abstract}

\newpage
\section{Introduction}\label{Intr}
In recent years there has been a renewed interest~\cite{continentino94,%
Belitz94,sachdev96,Sondhi96} in the theory of zero--temperature quantum
phase transitions initiated in 1976 by Hertz's
quantum dynamic renormalization group \cite{Hertz76} for itinerant
ferromagnets. Distinctively from temperature driven critical phenomena,
these phase transitions occur at zero temperature as a function of some
non--thermal control parameter (or a competition between different
parameters describing the basic interaction of the system), and the relevant
fluctuations are of quantum rather than thermal nature.

It is well known from the theory of critical phenomena that for the
temperature driven phase transitions quantum effects are unimportant near
critical points with $T_{c}>0$. It could be expected, however, that at
rather small (as compared to characteristic excitation in the system)
temperature, the leading $T$ dependence of all observables is specified by
the properties of the zero--temperature critical points, which take place in
quantum systems. The dimensional crossover rule asserts that the critical
singularities of such a quantum system at $T=0$ with dimensionality $d$ are
formally equivalent to those of a classical system with dimensionality $d+z$
($z$ is the dynamical critical exponent) and critical temperature $T_{c}>0$.
This makes it possible to investigate low--temperature effects (considering
an effective system with $d$ infinite space and $z$ finite time dimensions)
in the framework of the theory of finite--size scaling (FSS). The idea of
this theory has been applied to explore the low--temperature regime in
quantum systems (see Refs. \onlinecite{%
chakravarty89,sachdev92,chubukov94,sachdev94}), when the properties of the
thermodynamic observables in the {\it finite--temperature quantum critical
region} have been the main focus of interest. The very {\it quantum critical
region} was introduced and studied first by Chakravarty et al~\cite
{chakravarty89} using the renormalization group methods. The most famous
model for discussing these properties is the quantum nonlinear ${\cal O}(n)$
sigma model (QNL$\sigma$M).\cite
{chakravarty89,sachdev92,chubukov94,sachdev94,Rosenstein90,neuberger89,fisher89,hasenfratz93,azaria93,Castro93,fujii95}

Recently an equivalence between the QNL$\sigma$M in the limit $%
n\rightarrow\infty$ and a quantum version of the spherical model or more
precisely the ``spherical quantum rotors'' model (SQRM) was announced.\cite
{vojta95} The SQRM is an interesting model on its own. Due to the remarkable
opportunity it offers for a rigorous study of finite--size effects at
arbitrary dimensionality SQRM may play the same role in quantum critical
phenomena as the popular Berlin--Kac spherical model in classical critical
phenomena. The last one became a touchstone for various scaling hypotheses
and a source of new ideas in the general theory of finite size scaling (see
for example Refs.~\onlinecite{singh85,brankov88,allen89,privman90,singh92,%
danchev93,brankov94,allen94,allen95,chamati96} and references therein). Let
us note that an increasing interest related with the spherical approximation
(or large $n$--limit) generating tractable models in quantum critical
phenomena has been observed in the last few years.\cite{vojta95,tu94,nieu95,vojta96}

In Ref.~\onlinecite{vojta95}, the critical exponents for the
zero--temperature quantum fixed point and the finite--temperature classical
one as a function of dimensionality was obtained. What remains beyond the
scope of Ref.~\onlinecite{vojta95} is to study in an exact manner the
scaling properties of the model in different regions of the phase diagram
including the {\it quantum critical region} as a function of the
dimensionality of the system. In the context of the finite--size scaling
theory both cases: ({\bbox i}) The infinite $d$--dimensional quantum system
at low--temperatures $\infty ^d\times \!\!L_\tau ^z$ ($L_\tau \sim \left( 
\frac \hbar {k_B^{}T}\right) ^{1/z}$ is the finite--size in the imaginary
time direction) and ({\bbox{ii}}) the finite system confined to the geometry 
$L^{d-d^{\prime }}\!\!\times \!\!\infty ^{d^{\prime }}\!\!\times \!\!L_\tau
^z$ ($L$--linear space size) are of crucial interest.

In this paper a detailed theory of the scaling properties of the quantum
spherical model with nearest--neighbor interaction is presented. The
plan of the paper is as follows: we start with a brief review of the model
and the basic equations for the free energy and the quantum spherical field
in the case of periodic boundary conditions (Section~\ref{model}). The
relation of this model with the model due to Schneider, Stoll and Beck\cite
{plakida86,tonchev91,verbeure92,pisanova93,chamati94,pisanova95} has
been briefly commented in Section~\ref{other}. Since we would like to
exploit the ideas of the FSS theory, the bulk system in the low--temperature
region is considered like an effective ($d+1$)--dimensional classical system
with one finite (temporal) dimension. This is done to enable contact to be
made with other results based on the spherical type approximation e.g. in
the framework of the spherical model and the QNL$\sigma$M in the limit $%
n\rightarrow\infty$. The scaling forms of the free energy and of the
spherical field equation are derived for the infinite
(Section~\ref{ISYS}) and finite (Section~\ref{FSYS}) system confined to
the general geometry $L^{d-d^{\prime}}\!\!\times\!\!\infty^{d^{\prime}}
\!\!\times\!\! L_{\tau}^{z}$. In Section~\ref{analysis} we analyze in
detail the equation for the spherical field. This equation turns out
to allow for analytic studies of the finite--size and low--temperature
asymptotes for different $d$ and $d^{\prime}$. Special attention is
laid on the two dimensional system. The remainder of the paper contains
the details of the calculations: Appendices~\ref{app2} and~\ref{app3}.

\section{The Model}\label{model}
The model we will consider here describes a magnetic ordering
due to the interaction of quantum spins. This has the following
form\cite{vojta95}
\begin{equation}
{\cal H}=\frac 12g\sum_\ell {\cal P}_\ell ^2-\frac 12\sum_{\ell \ell
^{\prime }}{\bbox J}_{\ell \ell ^{\prime }}^{}{\cal S}_\ell ^{}{\cal S}%
_{\ell ^{\prime }}^{}+\frac \mu 2\sum_\ell {\cal S}_\ell ^2-H\sum_\ell {\cal
S}_\ell ^{},  \label{model1}
\end{equation}
where ${\cal S}_\ell ^{}$ are spin operators at site $\ell $, the operators
${\cal P}_\ell ^{}$ are ``conjugated'' momenta (i.e. $[{\cal S}_\ell ^{},%
{\cal S}_{\ell ^{\prime }}^{}]=0$, $[{\cal P}_\ell ^{},{\cal P}_{\ell
^{\prime }}^{}]=0$, and $[{\cal P}_\ell ^{},{\cal S}_{\ell ^{\prime
}}^{}]=i\delta _{\ell \ell ^{\prime }}^{}$, with $\hbar =1$), the coupling
constants ${\bbox J}_{\ell ,\ell ^{\prime }}^{}={\bbox J}$ are between
nearest neighbors only,~\cite{dyn} the coupling constant $g$ is introduced
so as to measure the strength of the quantum fluctuations (below it will be
called quantum parameter), $H$ is an ordering magnetic field, and finally
the spherical field $\mu $ is introduced so as to ensure the constraint
\begin{equation}
\sum_\ell \left\langle {\cal S}_\ell ^2\right\rangle =N.  \label{model2}
\end{equation}
Here $N$ is the total number of the quantum spins located at sites
``$\ell $'' of a finite hypercubical lattice $\Lambda $ of size
$L_1\times L_2\times\cdots \times L_d=N$ and $\left\langle\cdots\right\rangle$
denotes the standard thermodynamic average taken with ${\cal H}$.

Let us note that the commutation relations for the operators ${\cal S}_\ell
^{}$ and ${\cal P}_\ell ^{}$ together with the quadratic kinetic term in the
Hamiltonian~(\ref{model1}) do not describe quantum Heisenberg--Dirac spins
but quantum rotors as it was pointed out in Ref.~\onlinecite{vojta95}.

The free energy of the model in a finite region $\Lambda $ is given by the
Legendre transformation 
\begin{equation}
f_\Lambda \left( \beta ,g,H\right) :=\sup_\mu \left\{ -\frac 1{N\beta }\ln
Z_\Lambda \left( \beta ,g,H;\mu \right) -\frac 12\mu \right\} ,  \label{fed}
\end{equation}
where $Z_\Lambda \left( \beta ,g,H;\mu \right)=\text{Tr}\left[ \exp \left(
-\beta {\cal H}\right) \right] $ is the partition function of the model
and $\beta$ is the inverse temperature with the Boltzmann constant $K_B=1$.
Under periodic boundary conditions applied across the finite dimensions the
free energy takes the form 
\begin{equation}
\beta f_\Lambda \left( \beta ,g,H\right) =\sup_\mu \left\{ \frac 1N\sum_{
\bbox q}\ln \left[ 2\sinh \left( \frac 12\beta \omega \left( \bbox q;\mu
\right) \right) \right] -\frac{\beta g{\bbox J}}{2\omega ^2\left( \bbox
0;\mu \right) }H^2\right\} .  \label{fec}
\end{equation}
Here the vector $\bbox q$ is a collective symbol, which has for $L_j$ odd
integers, the components:
$$
\left\{ \frac{2\pi n_1}{L_1},\cdots ,\frac{2\pi n_d}{L_d}\right\} ,\ \
n_j\in \left\{ -\frac{L_j-1}2,\cdots ,\frac{L_j-1}2\right\} , 
$$
and 
$$
\omega^2 \left( \bbox q;\mu \right) =g\left( \mu -2{\bbox J}\sum_{i=1}^d\cos
q_i\right) . 
$$
The supremum of the free energy is attained at the solutions of the
mean--spherical constraint, Eq. (\ref{model2}), that reads
\begin{mathletters}\label{model3}
\begin{equation}
1=\frac \lambda {2N}\sum_{\bbox q}\frac 1{\sqrt{\phi +2\sum_{i=1}^d(1-\cos
q_i)}}\coth \left( \frac \lambda {2t}\sqrt{\phi +2\sum_{i=1}^d(1-\cos q_i)}%
\right) +\frac{h^2}{\phi ^2},  \label{model3a}
\end{equation}
which is equivalent to the following
\begin{equation}
1=\frac tN\sum_{m=-\infty }^\infty \sum_{\bbox q}\frac 1{\phi
+2\sum_{i=1}^d(1-\cos q_i)+b^2m^2}+\frac{h^2}{\phi ^2},  \label{model3b}
\end{equation}
\end{mathletters}
where we have introduced the notations: $\lambda =\frac g{\sqrt{\bbox J}}$
is the normalized quantum parameter, $t=\frac T{\bbox
J}$--the normalized temperature, $h=\frac H{\sqrt{\bbox J}}$ -- the
normalized magnetic field, $b=\frac{2\pi t}\lambda $, and $\phi =\frac \mu {%
\bbox J}-2d$ is the shifted spherical field.

Eqs.~(\ref{fec}) and (\ref{model3}) provide the basis of the study of the
critical behavior of the model under consideration.

A previous direct analysis~\cite{vojta95} of Eq.~(\ref{model3}) in the
thermodynamic limit shows that there can be no long--range order at finite
temperature for $d\leq 2$ (in accordance with the Mermin--Wagner theorem).
For $d>2$ one can find long--range order at finite temperature up to a
critical temperature $t_c(\lambda )$. Here we shall consider the
low--temperature region for $1<d<3$. Before passing to the corresponding
analysis we would like to make some comments on the relations of the model
considered here with other models known in the literature.

\section{Relations to other models}\label{other}
Recently\cite{plakida86,tonchev91,verbeure92,pisanova93,chamati94,pisanova95}
another model, suitable to handle the joint description of classical and
quantum fluctuations in an exact manner, was considered. If we consider
${\cal P}_\ell$ and ${\cal S}_\ell$ as canonically conjugated momentum and
coordinate, respectively, of the $\ell$ atom with mass $g^{-1}$ at each
lattice point $\ell\in{\Bbb Z}^d$, then the first three terms in
Eq.~(\ref{model1}) describe a harmonic interaction of the $\ell$--th
and $\ell'$--th atoms. One can see that in the absence of a spherical
constraint, if $A=\frac{\mu}{4}+\frac{d\bbox J}{2} <0$, such a lattice is
unstable i.e. the parameter $A<0$ defines the frequency of a mode
unstable in the harmonic approximation suggesting that an appropriate
stabilization of the lattice can create a gap in the phonon spectrum. In
the spirit of the self--consistent phonon approximation~\cite{plakida86}, it is
possible to add the term
\begin{equation}
\frac{B}{4N}\left(\sum_\ell{\cal S}_\ell^2\right)^2, \ \ \ \ \ B>0,
\end{equation}
``switching on'' an anharmonic interaction which stabilizes the lattice.
Because the term is inversely proportional to the particle number $N$
the model under consideration turns out to be exactly soluble in the
thermodynamic limit.~\cite{plakida86,pisanova93}

In normal coordinates the ``new'' model Hamiltonian reads
\begin{equation}\label{anh}
{\cal H}_{\text{anh.}}=\frac{1}{2}\sum_{\bbox q}\left({\cal P}_{\bbox q}
{\cal P}_{-\bbox q}+\omega_{h}^2({\bbox q}){\cal S}_{\bbox q}{\cal S}_{-\bbox q}
\right)+\frac{b}{4N}\sum_{\bbox{qq'}}{\cal S}_{\bbox q}{\cal S}_{-\bbox q}
{\cal S}_{\bbox q'}{\cal S}_{-\bbox q'}
\end{equation}
where ${\cal P}_{\bbox q}$ and ${\cal S}_{\bbox q}$ are Fourier components of
the operators ${\cal P}_\ell$ and ${\cal S}_\ell$, respectively. The
frequency of phonons is given by
\begin{equation}
\omega_{h}^2({\bbox q})=-\nu_0^2+2g{\bbox J}
\sum_{i=1}^d\left(1-\cos q_i\right).
\end{equation}
The anharmonicity constant $b=g^2B=\frac{\nu^4_0}{4E_0}$, where
$E_0=\frac{A^2}{4B}$ is the barrier height of the double--well potential
in~(\ref{anh}) at a uniform displacement of all particles:
${\cal S}_\ell\to x$;
$$
U(x)=-\frac{A}{2}x^2+\frac{B}{4}x^4.
$$

In Ref.~\onlinecite{plakida86} the following pseudoharmonic approximating
Hamiltonian has been proposed
\begin{equation}\label{appr}
{\cal H}_{\text{appr.}}(\Delta)=\frac{1}{2}\sum_{\bbox q}\left({\cal P}_{\bbox q}
{\cal P}_{-\bbox q}+\Omega_{h}^2({\bbox q},\Delta){\cal S}_{\bbox q}{\cal S}_{-\bbox q}
\right)-NE_0(1+\Delta)^2,
\end{equation}
where the trial harmonic frequency
\begin{equation}
\Omega^2_{h}({\bbox q},\Delta)=\omega^2_{h}({\bbox q})+\nu_0^2(1+\Delta)
\end{equation}
is defined by the equation for the temperature--dependent gap $\Delta$ (c.f.
with Eq.~(\ref{model3}))  :
\begin{equation}\label{gap}
1+\Delta=\frac{b}{N\nu^2_0}\sum_{\bbox q}\left<{\cal S}_{\bbox q}
{\cal S}_{-\bbox q}\right>_{{\cal H}_{\text{appr.}}}=
\frac{b}{2N\nu^2_0}\sum_{\bbox q}\frac1{\Omega_{anh}({\bbox q},\Delta)}
\coth\frac{\Omega_{anh}({\bbox q},\Delta)}{2T}.
\end{equation}

Eq.~(\ref{gap}) plays the role of the ``soft spherical constraint''. If
$B$ goes to zero, the quartic self--interaction disappears and our
model becomes a pure harmonic model. One can see that this limit is
singular and that $B$, for $d$ more than the upper critical
dimension, will be a dangerous irrelevant variable.~\cite{chamati94} The
model~(\ref{anh}) is a quantum counterpart of the ``soft'' classical mean
spherical model studied in Ref.~\onlinecite{shapiro86} in the context of
FSS theory.

The inspection of Eq.~(\ref{gap}) shows that it is similar to
Eqs.~(\ref{model3}) up to a linear term in the l.h.s of (\ref{gap})
(This term appears to be essential only above the upper critical
dimension.) Thus there is a clear mathematical analogy between the model
defined in Section~\ref{model} and the model presented here.

One of the main problems in studying the FSS effects for this type of
models is to answer the question: how correct is the finite--size
description of the initial model~(\ref{anh}) on the base of the
approximating Hamiltonian~(\ref{appr}). Evidently the answer to this
question will clarify the more general and subtle problem of the status
of finite--size results obtained by approximating methods. A successful
step in this direction is presented in Ref.~\onlinecite{brankov90} when the
classical mean spherical model in the Husimi--Temperely limit has been
considered. In Ref.~\onlinecite{brankov90} an appropriate modification
of the approximating Hamiltonian approach, which reproduced some exact
FSS results, has been suggested. Let us note that, if we parallel
Ref.~\onlinecite{brankov90}, for studying quantum systems  a more refined
mathematical problems must be solved and that up to now this is an open
problem.

\section{Scaling form of the free energy at low temperatures}\label{FEN}
Let us denote by $\tilde f_\Lambda \left( \beta ,g,H\right) $ the reduced
free energy $f_\Lambda \left( \beta ,g,H\right) /\bbox J$ . Then, from
Eq. (\ref{fec}) one obtains immediately
\begin{eqnarray}\label{fel}
\tilde f_\Lambda \left( t,\lambda ,h\right)  &=&\sup_\phi \left\{ \frac tN
\sum_{\bbox q}\ln \left[ 2\sinh \left( \frac \lambda {2t}\sqrt{\phi
+2\sum_{i=1}^d(1-\cos q_i)}\right) \right] \right.    \nonumber\\
& &-\left. \frac 12\frac{h^2}\phi -\frac 12\phi \right\} -d.  
\end{eqnarray}
Using the identities 
\begin{equation}
\ln \frac{\sinh b}{\sinh a}=\frac 12\sum_{m=-\infty }^\infty \ln \frac{%
b^2+\pi ^2m^2}{a^2+\pi ^2m^2},  \label{id1}
\end{equation}
where $ab>0$, $a,b$ are arbitrary real numbers and
\begin{equation}
\ln \left( a+b\right) =\ln a+\int_0^\infty \exp \left( -ax\right) \left(
1-\exp \left( -bx\right) \right) \frac{dx}x,  \label{id2}
\end{equation}
where $a>0,$ $a+b>0$ the above expression for the free energy can be
rewritten in the form
\begin{eqnarray}\label{f1}
\tilde f_\Lambda \left( t,\lambda ,h\right)  &=&\frac t2\sum_{m=-\infty
}^\infty \left\{ \int_0^\infty \frac{dx}x\exp \left[ -x\left( \phi
+b^2m^2\right) \right] \left[ 1-\frac 1{L_1}S_{L_1}(x)\times \ldots \times 
\frac 1{L_d}S_{L_d}(x)\right] \right\}  \nonumber\\
&&+ t\ln \left[ 2\sinh \left( \frac \lambda {2t}\sqrt{\phi }\right)
\right] -\frac 12\frac{h^2}\phi -\frac 12\phi -d.  
\end{eqnarray}
Here 
\begin{equation}
S_L(x)=\sum_{m=0}^{L-1}\exp \left[ -2x\left( 1-\cos \frac{2\pi m}L\right)
\right]   \label{SL}
\end{equation}
and $\phi $ in (\ref{f1}) is the solution of the corresponding spherical
field equation~(\ref{model3}). In the remainder of this article we will be
interested only in systems with geometry $L^{d-d^{\prime }}\times
\infty^{d^{\prime }}\times L_\tau$. With the help of the Jacobi identity
\begin{equation}\label{jacobi}
\sum_{m=-\infty}^{\infty} \exp\left(-um^{2}\right)=
\left(\frac{\pi}{u}\right)^{1/2}\sum_{m=-\infty}^{\infty}
\exp\left(-m^{2}\frac{\pi^{2}}{u}\right),
\end{equation}
the above expression for the free energy can
be rewritten in the following equivalent form 
\begin{eqnarray}
2\tilde f_\Lambda \left( t,\lambda ,h\right)  &=&\frac \lambda {\sqrt{4\pi }}%
\int_0^\infty \frac{dx}{x^{3/2}}\exp \left( -x\phi \right) \left\{ \left[
1-\left( \frac 1LS_L(x)\right) ^{d-d^{\prime }}\right] \left[ \exp \left(
-2x\right) I_0(2x)\right] ^{d^{\prime }}\right\} \nonumber \\
&&+ \frac \lambda{\sqrt{\pi}}\int_0^\infty \frac{dx}{x^{3/2}}\exp
\left( -x\phi \right) R\left( \frac{\pi ^2}{xb^2}\right) \left\{ \left[
1-\left( \frac 1LS_L(x)\right) ^{d-d^{\prime }}\right]\right.\nonumber \\
&&\times \left[ \exp \left(-2x\right) I_0(2x)\right]^{d^{\prime }}\Biggr\}
+2t\ln \left[ 2\sinh \left( \frac \lambda {2t}\sqrt{\phi }\right)
\right] -(\phi +2d)-\frac{h^2}\phi ,  \label{f2}
\end{eqnarray}
where 
\begin{equation}
R(x)=\sum_{m=1}^\infty \exp \left( -xm^2\right) ,  \label{R}
\end{equation}
and $I_0(x)$ is a modified Bessel function.

Dividing the integrals from $0$ to $\infty $ in the r.h.s of (\ref{f2}) into
integral from $0$ to $L^2$ and from $L^2$ to $\infty ,$ after using for
these intervals the corresponding asymptotics of $S_L(x)$ \cite{danchev93}
\begin{equation}\label{est1}
S_L(x)=1+2R\left( \frac{4\pi ^2}{L^2}x\right) -Lv(x)+O(\exp (-x\;const))
\label{s1}
\end{equation}
for $x>L^2,$ where 
$$
v(x)=\left( 4\pi x\right) ^{-1/2}\left[ 1-\mathop{\rm erf}
\left( \pi \sqrt{x}\right) \right] ,
$$
and 
\begin{equation}\label{est2}
S_L(x)=L\exp \left( -2x\right) I_0(2x)+\left( L/\sqrt{\pi x}\right) R\left( 
\frac{L^2}{4x}\right)   \label{s2}
\end{equation}
for $0<x\leq L^2,$ one ends up with the following expression for the free
energy at low temperatures ($\frac \lambda t\gg 1$) 
\begin{eqnarray}\label{f3}
2\tilde f_\Lambda \left( t,\lambda ,h\right)  &=&\lambda f_d\left( \phi
\right) +\lambda \sqrt{\phi}-(\phi+2d)-\frac{h^2}\phi \nonumber\\
&&+ \lambda \int_0^\infty \frac{dx}x\left( 4\pi x\right) ^{-(d+1)/2}\exp
\left( -x\phi \right) \nonumber\\
&&\times\left[ 1-\left( 1+2R\left( \frac{\pi ^2}{xb^2}\right)
\right) \left( 1+2R\left( \frac{L^2}{4x}\right) \right) ^{d-d^{\prime
}}\right] ,  
\end{eqnarray}
where 
\begin{equation}
f_d\left( \phi \right) =\frac 1{\sqrt{4\pi }}\int_0^\infty \frac{dx}{x^{3/2}}%
\exp \left( -x\phi \right) \left[ 1-\left( \exp \left( -2x\right)
I_0(2x)\right) ^d\right] .  \label{fd}
\end{equation}
The above expressions are valid for any $d.$ For $1<d<3$ it is easy to show
that 
\begin{equation}
f_d\left( \phi \right) =f_d\left( 0\right) +{\cal W}_d(0)\phi -\frac{4\Gamma
\left( \frac{3-d}2\right) }{\left( d^2-1\right) \left( 4\pi \right)
^{(d+1)/2}}\phi ^{(d+1)/2}-\sqrt{\phi },  \label{fdf}
\end{equation}
where 
\begin{equation}
{\cal W}_d(\phi )=\frac 1{2(2\pi )^d}\int_{-\pi }^\pi dq_1\cdots \int_{-\pi
}^\pi dq_d\left( \phi +2\sum_{i=1}^d(1-\cos q_i)\right) ^{-1/2}.
\label{bulk4}
\end{equation}
is the well--known Watson type integral.

Combining (\ref{f3}) and (\ref{fdf}) and denoting $\lambda_c:=1/{\cal W}_d(0)$
we obtain
\begin{eqnarray}\label{ff}
2\tilde f_L\left( t,\lambda ,h\right)  &=&\lambda f_d\left( 0\right)
-2d+\left( \frac \lambda {\lambda _c}-1\right) \phi -\frac{4\Gamma \left( 
\frac{3-d}2\right) \lambda }{\left( d^2-1\right) \left( 4\pi \right)
^{(d+1)/2}}\phi ^{(d+1)/2}-\frac{h^2}\phi \nonumber\\
&& +\lambda \int_0^\infty \frac{dx}x\left( 4\pi x\right) ^{-(d+1)/2}\exp
\left( -x\phi \right)\nonumber\\
&&\times\left[ 1-\left( 1+2R\left( \frac{\pi ^2}{xb^2}\right)
\right) \left( 1+2R\left( \frac{L^2}{4x}\right) \right) ^{d-d^{\prime
}}\right] .
\end{eqnarray}

\subsection{The infinite system}\label{ISYS}
Here we consider the behavior of the free energy of the bulk system for low
temperatures. From Eq. (\ref{ff}), taking the limit $L\rightarrow \infty $
one derives immediately
\begin{eqnarray}\label{ffb}
2\tilde f_\infty \left( t,\lambda ,h\right)  &=&\lambda f_d\left( 0\right)
-2d+\left( \frac \lambda {\lambda _c}-1\right) \phi -\frac{4\Gamma \left( 
\frac{3-d}2\right) \lambda }{\left( d^2-1\right) \left( 4\pi \right)
^{(d+1)/2}}\phi ^{(d+1)/2} \nonumber\\
&&-\frac{h^2}\phi -2\lambda \int_0^\infty \frac{dx}x\left( 4\pi x\right)
^{-(d+1)/2}\exp \left( -x\phi \right) R\left( \frac{\pi^2
}{xb^2}\right) ,
\end{eqnarray}
where $\phi$ is the solution of the corresponding equation for the bulk
system (see Eq. (\ref{bulk7}) below). Formally, one can obtain the
corresponding equation by requiring the first derivative, with respect to
$\phi$, of the r.h.s of (\ref{ffb}) to be zero. Therefore, the singular (the
$\phi $ dependent) part of the free energy can be rewritten in the form
\begin{equation}
\tilde f_{sing,bulk}\left( t,\lambda ,h\right) =\frac 12\lambda
b^{d+1}X_b\left(x_\lambda ,h_\lambda \right) ,  \label{sf1}
\end{equation}
where
\begin{equation}
x_\lambda =\left( \frac 1\lambda -\frac 1{\lambda _c}\right) b^{-(d-1)}
\label{xl}
\end{equation}
and
\begin{equation}
h_\lambda =hb^{-(d+3)/2}\lambda ^{-1/2}  \label{hl}
\end{equation}
are the scaling variables and 
\begin{eqnarray}\label{Xin}
X_b\left( x_\lambda ,h_\lambda \right)  &=&-x_\lambda y_\lambda -\frac{%
4\Gamma \left( \frac{3-d}2\right) }{\left( d^2-1\right) \left( 4\pi \right)
^{(d+1)/2}}y_\lambda ^{(d+1)/2}-\frac{h_\lambda ^2}{y_\lambda }\nonumber\\
&&-2\int_0^\infty \frac{dx}x\left( 4\pi x\right) ^{-(d+1)/2}\exp \left(
-xy_\lambda \right) R\left(\frac{\pi^2}x\right)
\end{eqnarray}
is the universal scaling function for the free energy. In the last expression
$y_\lambda :=\phi/b^2$ is the solution of the spherical field equation
\begin{equation}\label{sfeb}
0 =-x_\lambda +\frac{\Gamma \left( \frac{1-d}2\right) }{\left( 4\pi \right) ^{(d+1)/2}}y_\lambda ^{(d-1)/2}+\frac{%
h_\lambda ^2}{y_\lambda ^2}
-\frac 1{2\pi }\int_0^\infty \frac{dx}x\left( 4\pi x\right) ^{-(d-1)/2}\exp
\left( -xy_\lambda \right) R\left( \frac{\pi ^2}x\right).
\end{equation}

\subsection{The finite system}\label{FSYS}
For the finite system from Eq. (\ref{ff}), in a full analogy with the
previous results for the infinite system, we derive
\begin{equation}
\tilde f_{sing,finite}\left( t,\lambda ,h,L\right) =\frac 12\lambda
b^{d+1}X_{finite}\left( x_\lambda ,h_\lambda ,a\right)   \label{sf2}
\end{equation}
where the new scaling variable 
\begin{equation}
a=Lb  \label{a}
\end{equation}
is introduced, whereas the scaling function now is 
\begin{eqnarray}\label{Xfin}
X_{finite}\left( x_\lambda ,h_\lambda ,a\right)  &=&-x_\lambda y_\lambda -
\frac{4\Gamma \left( \frac{3-d}2\right) }{\left( d^2-1\right) \left( 4\pi
\right) ^{(d+1)/2}}y_\lambda ^{(d+1)/2}-\frac{h_\lambda ^2}{y_\lambda }
\nonumber \\
&&+\int_0^\infty \frac{dx}x\left( 4\pi x\right) ^{-(d+1)/2}\exp \left(
-xy_\lambda \right) \nonumber\\
&&\times\left[ 1-\left( 1+2R\left(\frac{\pi ^2}x\right) \right)
\left( 1+2R\left( \frac{a^2}{4x}\right) \right) ^{d-d^{\prime }}\right] . 
\end{eqnarray}
Here $y_\lambda $ is the solution of the corresponding spherical field
equation for the finite system which reads
\begin{eqnarray}\label{sfef}
0 &=&-x_\lambda +\frac{\Gamma\left(\frac{1-d}2\right)}{\left(4\pi\right)
^{(d+1)/2}}y_\lambda ^{(d-1)/2}+\frac{%
h_\lambda ^2}{y_\lambda ^2}
+\frac 1{4\pi }\int_0^\infty \frac{dx}x\left( 4\pi x\right) ^{-(d-1)/2}\exp
\left( -xy_\lambda \right) \nonumber\\
&&\times\left[ 1-\left( 1+2R\left( \frac{\pi ^2}x\right)
\right) \left( 1+2R\left( \frac{a^2}{4x}\right) \right) ^{d-d^{\prime
}}\right] .
\end{eqnarray}

The expressions (\ref{Xin}) and (\ref{sfeb}) for the infinite system, and
(\ref{Xfin}) and (\ref{sfef}) for the finite one, represent actually the
verification of the analog of the Privman--Fisher
hypothesis~\cite{fisher84} for the finite--size scaling form of the free
energy, formulated  for classical systems, for the
case when the quantum fluctuations are essential. Note, that in that case
one has both finite space dimensions and one additional finite dimension
that is proportional to the inverse temperature. According to the
finite--size scaling hypothesis~\cite{sachdev94,fisher84} one has to expect that the
temperature dependent scaling field multiplying the universal scaling
function will be with exponent $p=1+d/z$, where the dynamic--critical
exponent $z$ expresses the anisotropic scaling between space and
``temperature'' (``imaginary--time'') directions. For the system considered
here $z=1$, in full conformity with our results. It seems important enough
to emphasize that in the quantum low--temperature case there is one nonuniversal
pre-factor (for the system considered here it is $J\lambda$) that multiplies
the universal finite--size scaling function of the free energy, which is not
the case for the classical systems (provided their free--energy density is
normalized per $k_BT$).\cite{fisher84}

Now we pass to the analysis of the spherical field equations for both the
finite and infinite systems following Ref.~\cite{chamati97}

\section{Analysis of the spherical field equations}\label{analysis}
\subsection{The infinite system}\label{bulk}
After performing the integration in the right--hand side of the equation for
the spherical field~(\ref{sfeb}) and using the integral
representation~(\ref{IR}) we get
\begin{equation}  \label{bulk7}
\frac{1}{\lambda}-\frac{1}{\lambda_{c}}=-\frac{1}{(4\pi)^{(d+1)/2}}
\left|\Gamma \left( \frac{1-d}{2} \right) \right|\phi^{(d-1)/2} + \frac{2}{%
(4\pi)^{(d+1)/2}}\phi^{(d-1)/2} {\cal K} \left(\frac{d-1}{2},\frac{\lambda}{%
2t}\phi^{1/2}\right) + \frac{h^{2}}{\phi^{2}},
\end{equation}
where 
\begin{equation}
{\cal K}(\nu,y) = 2 \sum_{m=1}^{\infty} (y m)^{-\nu} K_{\nu} (2 m y)
\end{equation}
and $K_\nu(x)$ is the other modified Bessel function.

Eq.~(\ref{bulk7}) was derived in Ref.~\onlinecite{chamati97} by a
different technique. The approach used here is based on the
estimations~(\ref{est1},\ref{est2}) and has the advantage that it can
easily be generalized for other boundary conditions.\cite{danchev93}

In the remainder of this section we will study the effect of the temperature
on the susceptibility and the equation of state near the quantum critical
fixed point.


\subsubsection{Zero--field susceptibility}\label{ZFSUS}

\begin{figure}[tbp]
\epsfxsize=4.5in
\centerline{\epsffile{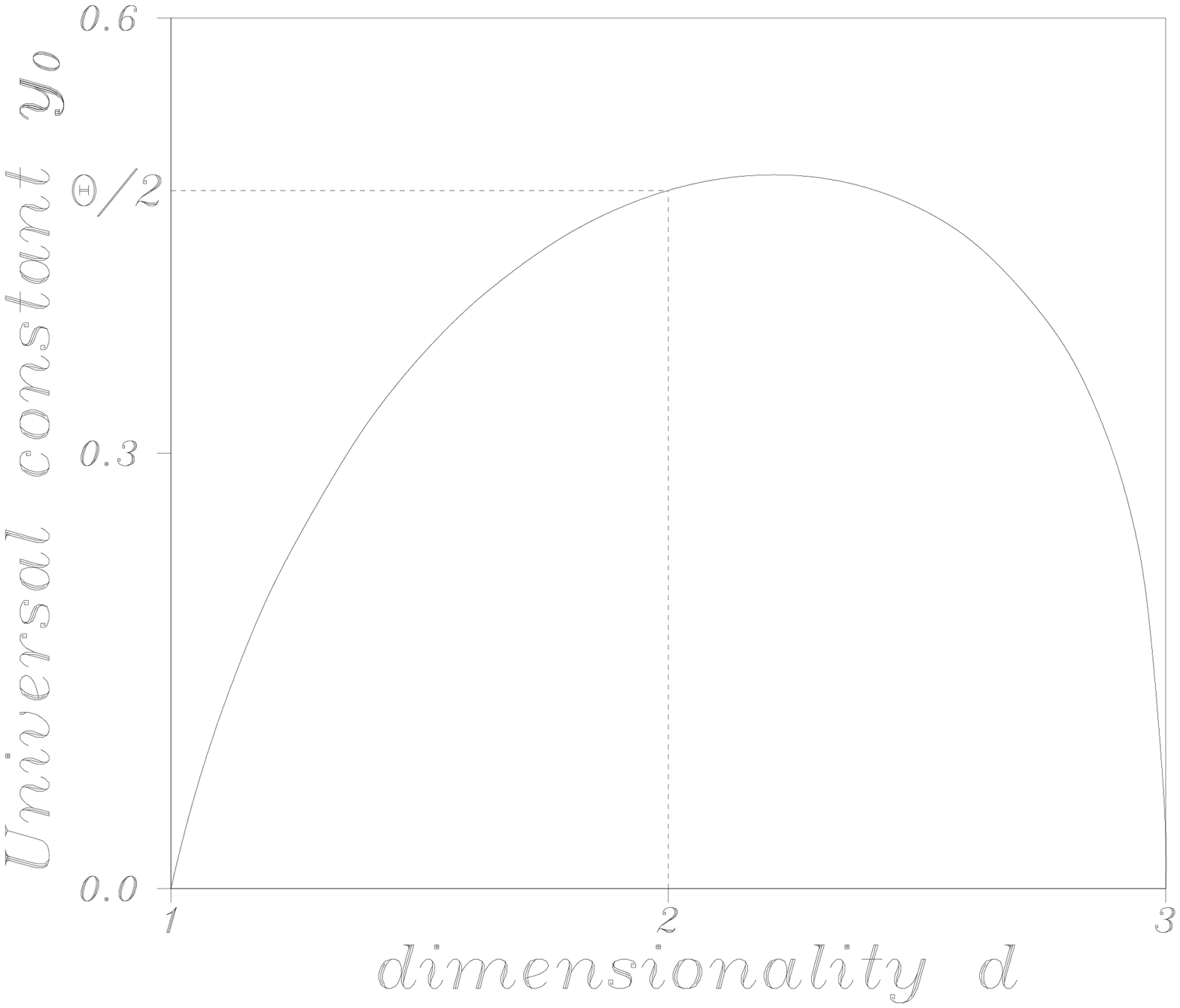}}
\vspace{0.1in}
\caption{The dependence of the universal constant $y_{0}$ upon the
dimensionality $d$. The constant $\Theta=0.962424...$ is obtained for the
two--dimensional system (see Eq.~(\protect\ref{theta}))}
\label{fig1}
\end{figure}

After making the field $h$ vanishes, from Eq.~(\ref{bulk7}) we find that the
normalized zero--field susceptibility $\chi=\phi^{-1}$ on the line $\lambda
= \lambda_{c}$ ($t\rightarrow0^{+}$) is given by 
\begin{equation}  \label{bulk8}
\chi = \frac{\lambda_{c}^{2}}{4 y_{0}^{2}} t^{-2},
\end{equation}
where $y_{0}$ is the universal solution of 
\begin{equation}  \label{scal1}
\left|\Gamma \left( \frac{1-d}{2} \right) \right| = 2 {\cal K}\left(\frac{d-1%
}{2},y\right).
\end{equation}
The behavior of the universal constant $y_{0}$ as a function of the
dimensionality $d$ of the system is shown in FIG.~\ref{fig1}.

Eq.~(\ref{bulk8}) tells us that at low--temperature the
susceptibility increases as the inverse of the square of the temperature
above the quantum critical point.

In what follows we will try to investigate Eq.~(\ref{bulk7}) for different
dimensions ($1<d<3$) of the system and in different regions of the ($%
t,\lambda$) phase diagram. To this end we will consider first the function $%
{\cal K}(\nu,y)$. For $d\neq 2$ its asymptotic form is (for $y \ll 1$)~\cite
{singh89} (see also Appendix~\ref{app2}) 
\begin{eqnarray}
{\cal K}\left(\frac{d-1}{2},y\right)&\equiv&{\cal K}_{1}\left.\left( \frac{%
d-1}{2}\right|1,y\right)  \nonumber \\
&\approx& \frac{\pi^{1/2}}{2}\Gamma\left(1-\frac{d}{2}\right) y^{-1}+
\Gamma\left(\frac{d-1}{2}\right)\zeta(d-1)y^{1-d}-\frac{1}{2} \Gamma\left(%
\frac{1-d}{2}\right),  \label{bulk9}
\end{eqnarray}
where $\zeta (x)$ is the Riemann zeta function. Introducing the ``shifted''
critical value of the quantum parameter due to the temperature by 
\begin{mathletters}
\label{bulk10}
\begin{equation}
\frac{1}{\lambda_{c}(t)}=\frac{1}{\lambda_{c}}+\frac{1} {2\pi^{(d+1)/2}}%
\left(\frac{t}{\lambda_{c}(t)}\right)^{d-1} \Gamma\left(\frac{d-1}{2}%
\right)\zeta(d-1),
\end{equation}
or 
\begin{equation}
\frac{1}{\lambda_{c}^{\mp}(t)}\approx\frac{1}{\lambda_{c}} \mp\frac{1}{%
2\pi^{(d+1)/2}}\left(\frac{t}{\lambda_{c}}\right)^{d-1} \Gamma\left(\frac{d-1%
}{2}\right)|\zeta(d-1)|,
\end{equation}
one has to make a difference between the two cases $d<2$ ``sign $-$'' and $d>2$
``sign $+$''. In the first case ($1<d<2$), it is possible to define the {\it %
quantum critical region} by the inequality 
\end{mathletters}
\begin{equation}
\left|\frac{1}{\lambda}-\frac{1}{\lambda_{c}}\right|\ll \frac{1}{%
2\pi^{(d+1)/2}}\left(\frac{t}{\lambda_{c}(t)}\right)^{d-1} \Gamma\left(\frac{%
d-1}{2}\right)|\zeta(d-1)|.
\end{equation}

For $d<2$ the function ${\cal K}(\nu,y) \sim y^{-1}$, by substitution in~(
\ref{bulk7}) we obtain for $\lambda<\lambda_{c}$ (outside of the {\it
quantum critical region}) 
\begin{equation}  \label{suscepb}
\chi \approx \left[\frac{\left|\Gamma\left(1-\frac{d}{2}\right)\right|}{%
(4\pi) ^{d/2}}\frac{\lambda_{c}}{\lambda_{c}-\lambda}\right]^{2/(d-2)}
t^{2/(d-2)}.
\end{equation}
We see that the susceptibility is going to infinity with power law degree
when the quantum fluctuations become important ($t\rightarrow0^{+}$) and
there is no phase transition driven by $\lambda$ in the system for
dimensions between 1 and 2.

In the second case ($2<d<3$), after the insertion of Eq.~(\ref{bulk9}) in
Eq.~(\ref{bulk7}) one has
\begin{equation}  \label{bulk11}
\chi \approx \left[\frac{\left|\Gamma\left(1-\frac{d}{2}\right)\right|}{%
(4\pi) ^{d/2}}\frac{\lambda_{c}(t)}{\lambda-\lambda_{c}(t)}\right]^{2/(d-2)}
t^{2/(d-2)},
\end{equation}
as a solution for $\lambda$ less than $\lambda_{c}$ and greater than the
critical value $\lambda_{c}(t)$ of the quantum parameter. Here for finite
temperatures there is a phase transition driven by the quantum parameter $%
\lambda$ with critical exponent of the $d$--dimensional classical spherical
model $\gamma= \frac{2}{d-2}$. This however is valid only for very close
values of $\lambda$ to $\lambda(t)$. For $\lambda<\lambda_{c}(t)$ the
susceptibility is infinite.

In the region where $\lambda>\lambda_{c}$ the zero--field susceptibility is
given by 
\begin{equation}  \label{bulk12}
\chi\approx\left[\frac{(4\pi)^{(d+1)/2}}{\Gamma\left(\frac{1-d}{2} \right)}%
\left(\frac{1}{\lambda}-\frac{1}{\lambda_{c}}\right) \right]^{\frac{2}{1-d}}.
\end{equation}
This result is valid for every $d$ between the lower and the upper quantum
critical dimensions.

Eq.~(\ref{bulk9}) is derived for dimensionalities $d\neq 2$. The important
case $d=2$ requires special care. In this case Eq.~(\ref{bulk7}) takes the
form ($h=0$) 
$$
\frac{\lambda}{\lambda_{c}} - 1=\frac{\lambda\phi^{1/2}}{4\pi} + \frac{t}{%
2\pi} \ln\left[1-\exp\left(-\frac{\lambda}{t} \phi^{1/2}\right)\right]. 
$$
This equation can be solved easily and one gets 
\begin{equation}  \label{bulk13}
\phi^{1/2}=\frac{2t}{\lambda}\text{arcsinh}\left\{\frac{1}{2} \exp\left[%
\frac{2\pi}{t}\left(\frac{\lambda}{\lambda_{c}}-1\right) \right]\right\}
\end{equation}

For the susceptibility, Eq.~(\ref{bulk13}) yields 
\begin{mathletters}
\label{bulk14}
\begin{equation}
\chi \approx \frac{\lambda^{2}}{t^{2}} \exp\left(4\pi \frac{%
\lambda_{c}-\lambda}{t\lambda_{c}}\right)
\end{equation}
for $\frac{2\pi}{t}\left|\frac{\lambda}{\lambda_{c}}-1\right|\gg1$ and $%
\lambda < \lambda_{c}$ i.e. in the renormalized classical region. For $%
\lambda=\lambda_{c}=3.1114...$ 
\begin{equation}
\chi=\frac{1}{\Theta^{2}}\left(\frac{\lambda_{c}}{t}\right)^{2},
\end{equation}
where the universal constant 
\begin{equation}  \label{theta}
\Theta= 2y_{0}=2\ln\left(\frac{\sqrt{5}+1}{2}\right)= -2\ln\left(\frac{\sqrt{%
5}-1}{2}\right)=.962424...
\end{equation}
was first obtained in the framework of the 3-dimensional classical mean
spherical model with one finite dimension.\cite{singh85} Finally for $\frac{%
2\pi}{t}\left|\frac{\lambda}{\lambda_{c}}-1\right|\gg1$ and $\lambda >
\lambda_{c}$ i.e. in the quantum disordered region
\begin{equation}  \label{bulk14a}
\chi\approx\left[\frac{\lambda\lambda_{c}}{4\pi(\lambda-\lambda_{c})}
\right]^{2}\left\{1+\frac{2t\lambda_{c}}{\pi(\lambda-\lambda_{c})}
\exp\left[-\frac{4\pi}{t\lambda_{c}}(\lambda-\lambda_{c})\right] \right\}.
\end{equation}
The first term of Eq.~(\ref{bulk14a}) is a particular case of Eq.~(\ref
{bulk12}) for $d=2$.

From Eqs.~(\ref{bulk14}) one can transparently see the different behaviors
of $\chi(T)$ in tree regions: a) renormalized classical region with
exponentially divergence as $T\rightarrow 0$, b) {\it quantum critical
region} with $\chi(T)\sim T^{-2}$ and crossover lines $T\sim|\lambda-
\lambda_{c}|$, and c) quantum disordered region with temperature independent
susceptibility (up to exponentially small corrections) as $T\rightarrow 0$.
The above results~(\ref{bulk13}) and~(\ref{bulk14}) coincide in form with
those obtained in Refs.~\onlinecite{chubukov94,Rosenstein90} for the two
dimensional QNL$\sigma$M in the $n\rightarrow\infty$ limit. The only
differences are: in Eq.~(\ref{bulk13}) the temperature is scaled by $\lambda$,
and the critical value $\lambda_{c}$ is given by $\lambda_c=1/{\cal W}_d(0)$
while for the QNL$\sigma$M it depends upon the regularization scheme. It
will be useful to clarify the bulk critical behavior of $\chi$ found above
in the context of FSS theory. First, from Eqs.~(\ref{bulk10}) one can see
that the exponent $\psi$ characterizing the shift in the quantum parameter
$\lambda_{c}$ by the temperature $t$ is equal to $\nu^{-1}=d-1$ in accordance
with FSS prediction. Second, let us consider the critical behavior of ($d+1$%
)--dimensional classical model with $L_{\tau}\sim 1/T$ playing the role of a
finite--size in the imaginary time direction, i.e. with slab geometry $%
\infty^{d}\times L_{\tau}$. FSS calculation (see for example Ref.~%
\onlinecite{allen89}) for the susceptibility at the critical point (in the
case under consideration this is $\lambda=\lambda_{c}$) gives: ({\bbox i})
for $1<d<2$,\cite{index} 
\end{mathletters}
$$
\chi(\lambda_{c})\sim L_{\tau}^{\gamma_{\lambda}/\nu_{\lambda}}, 
$$
where $\gamma_{\lambda}=2\nu_{\lambda}=\frac{2}{d-1}$ and $%
\chi(\lambda_{c})=T^{-\gamma_{T}^{}}$; $\gamma_{T}^{}=2$, (\bbox{ii}) for $%
2<d<3$, 
$$
\chi(\lambda)\sim L_{\tau}^{(\gamma_{\lambda}-\dot{\gamma}%
_{\lambda})/\nu_{\lambda}} \left|\lambda-\lambda(L_{\tau})\right|^{-\dot{%
\gamma}_{\lambda}}, 
$$
where the exponent $\dot{\gamma}_{\lambda}=\frac{2}{d-2}$ is pertaining to
the infinite size (space) dimensions $d$. In the last case for the shift of $%
\lambda_{c}$ we take from Eqs.~(\ref{bulk10}) $\lambda_{c}-\lambda(L_{\tau})%
\sim t^{d-1}$ and again $\chi(\lambda_{c})\sim t^{-\gamma_{T}^{}}$. So the
different types of critical exponent depending on $\lambda$ and $T$ are
related by $\gamma_{T}^{}=\gamma_{\lambda}/\nu_{\lambda}$ or $%
\gamma_{T}^{}=\psi\gamma_{\lambda}$ ($\psi=1/\nu_{\lambda}$ is the crossover
exponent).

%

\subsubsection{Equation of state}\label{EQST}

The equation of state of the Hamiltonian~(\ref{model1}) near the
quantum critical point is obtained after substituting the shifted spherical
field $\phi$ by the magnetization ${\cal M}$ through the relation 
\begin{equation}  \label{mag}
{\cal M} =\frac{h}{\phi},
\end{equation}
in Eq.~(\ref{bulk7}), which allows us to write the equation of state in a
scaling form 
\begin{equation}  \label{bulk15}
1+\frac{\delta\lambda}{{\cal M}^{1/\beta}} = (4\pi)^{-(d+1)/2} \left[\frac{h}{%
{\cal M}^{\delta}}\right]^{1/\gamma}\left\{\left| \Gamma\left(\frac{1-d}{2}%
\right)\right|- 2{\cal K}\left(\frac{d-1}{2}, \frac{\lambda}{2}\left(\frac{%
{\cal M}^{\nu/\beta}}{t} \right)\left(\frac{h}{{\cal M}^{\delta}}%
\right)^{\beta}\right) \right\},
\end{equation}
where 
$$
\delta\lambda=\frac{1}{\lambda_{c}}-\frac{1}{\lambda}. 
$$
We conclude that near the quantum critical point Eq.~(\ref{bulk15}) may be
written in general forms as 
\begin{mathletters}
\label{bulk16}
\begin{equation}
h={\cal M}^{\delta} f_{h} (\delta\lambda {\cal M}^{-1/\beta},
(t/\lambda)^{1/\nu}{\cal M}^{-1/\beta}),
\end{equation}
or 
\begin{equation}
{\cal M}=\left(\frac{t}{\lambda}\right)^{-\beta/\nu} f_{{\cal M}}
(\delta\lambda {\cal M}^{-1/\beta},h{\cal M}^{-\delta}).
\end{equation}

In Eqs.~(\ref{bulk16}) $f_{h}(x,y)$ and $f_{{\cal M}}(x,y)$ are some scaling
functions, furthermore $\gamma=\frac{2}{d-1}$, $\nu=\frac{1}{d-1}$, $\beta=%
\frac{1}{2}$ and $\delta=\frac{d+3}{d-1}$ are the familiar bulk critical
exponents for the $(d+1)$--dimensional classical spherical model. Eqs.~(\ref
{bulk16}) are a direct verification of FSS hypothesis in conjunction with
classical to quantum critical dimensional crossover. They can be easily
transformed into the scaling form (Eq. (21)) obtained in Ref.~%
\onlinecite{vojta95}, however here they are verified for $1<d<3$ instead of $%
2<d<3$ (c.f. Ref.~\onlinecite{vojta95}), i.e. the non--critical case is
included.

Hereafter we will try to give an explicit expression of the scaling function 
$f_{h}(x,y)$ ($x\equiv\delta\lambda{\cal M}^{-2}$, $y\equiv(t/\lambda)^{d-1}%
{\cal M}^{-2}$) in the neighborhood of the quantum critical fixed point.
This may be performed, in the case $\frac{t}{\lambda}\sqrt{h/{\cal M}}\ll1$,
with the use of the asymptotic form from Eq.~(\ref{bulk9}) to get (for $%
d\neq2$), 
\end{mathletters}
\begin{equation}  \label{bulk17}
\delta\lambda+{\cal M}^{2} +\frac{t/\lambda}{(4\pi)^{d/2}}\Gamma \left(1-%
\frac{d}{2}\right)\left(\frac{h}{{\cal M}}\right)^{(d-2)/2} + \frac{1}{%
2\pi^{(d+1)/2}} \left(\frac{t}{\lambda} \right)^{d-1} \Gamma\left(\frac{d-1}{%
2}\right)\zeta(d-1)=0.
\end{equation}
From this equation we find for the scaling function the result 
\begin{equation}
f_{h}(x,y)=\left[\frac{(4\pi)^{d/2}}{\Gamma\left(1-\frac{d}{2} \right)}
y^{-\nu}\left(1+x+\frac{1}{2\pi^{(d+1)/2}} \Gamma\left( \frac{d-1}{2}%
\right)\zeta(d-1) y\right)\right]^{2/(d-2)}.
\end{equation}

For the special case $d=2$ Eq.~(\ref{bulk15}) reads 
\begin{equation}
\delta\lambda=\frac{h^{1/2}}{4\pi{\cal M}^{1/2}} + \frac{t}{2\pi\lambda}
\ln\left[1-\exp\left(-\frac{\lambda}{t} \frac{h^{1/2}}{{\cal M}^{1/2}}%
\right)\right] -{\cal M}^{2},
\end{equation}
which yields 
\begin{equation}
h={\cal M}^{5} f_{h}(x,y)
\end{equation}
where the scaling function is given by the expression 
\begin{equation}  \label{bulk18}
f_{h}(x,y)= 4y^{2}\left[\text{arcsinh}\frac{1}{2}\exp\left(2\pi \frac{1+x}{y}%
\right)\right]^{2}.
\end{equation}

At $x=0$ and $y\gg1$ (fixed low--temperature and $h\rightarrow0^{+}$) the
scaling function~(\ref{bulk18}) reduces to 
\begin{equation}
f_{h}(0,y)\approx y^{2} \exp \left(\frac{4\pi}{y}\right).
\end{equation}
In the region $x<-1$, and for $y\ll1$ (fixed weak field and $%
t\rightarrow0^{+}$) the corresponding scaling function is 
\begin{equation}
f_{h}(x,y)\approx y^{2} \exp\left(4\pi\frac{x+1}{y}\right),
\end{equation}
and in region $x>-1$ and $y\ll1$ we have 
\begin{equation}  \label{bulk19}
f_{h}(x,y)\approx16\pi^{2}(x+1)^{2}\left[1+\frac{y}{\pi(1+x)}\exp\left( -4\pi%
\frac{1+x}{y}\right)\right].
\end{equation}
This identifies the zero temperature ($y=0$) form of the scaling function~(%
\ref{bulk19}) with those of the 3-dimensional classical spherical model.


\subsection{System confined to a finite geometry}\label{fing}
When the model Hamiltonian~(\ref{model1}) is confined to the
general geometry $L^{d-d^{\prime}}\!\!\times\!\!\infty^{d^{\prime}}\!\!%
\times\!\!L_{\tau}$, with $0\leq d^{\prime}\leq d$, equation~(\ref{sfef})
of the spherical field $\phi$ takes the form (after some algebra)
\begin{eqnarray}
\frac{1}{\lambda}&=&\frac{1}{\lambda_{c}}- (4\pi)^{-(d+1)/2} \left| \Gamma
\left( \frac{1-d}{2} \right) \right|\phi^{(d-1)/2}  \nonumber \\
& &+\frac{\phi^{(d-1)/2}}{(2\pi)^{(d+1)/2}} {\sum_{m, {\bbox l}%
(d-d^{\prime})}}^{\hskip-4mm\prime} \ \ \frac{K_{(d-1)/2} \left[\phi^{1/2}
\left\{(\lambda m/t)^{2} + (L|{\bbox l}|)^{2}\right\}^{1/2} \right]}{%
\left[\phi^{1/2} \left\{(\lambda m/t)^{2} + (L|{\bbox l}|)^{2}\right\}^{1/2}%
\right]^{(d-1)/2}} +\frac{h^{2}}{\phi^{2}} ,  \label{finite}
\end{eqnarray}
where 
$$
|{\bbox l}| = \left(l_{1}^{2}+l_{2}^{2}+\cdots+l_{d-d^{\prime}}^{2}
\right)^{1/2} 
$$
and the primed summation indicates that the vector with components $%
m=l_{1}=l_{2}=\cdots=l_{d-d^{\prime}}=0$ is excluded.


\subsubsection{Shift of the critical quantum parameter}\label{scqp}

The finite--size scaling theory (for a review see Ref.~\onlinecite{barber83}%
) asserts, for the temperature driven phase transition, that the phase
transition occurring in the system at the thermodynamic limit persists, if
the dimension $d^{\prime}$ of infinite sizes is greater than the lower
critical dimension of the system. In this case the value of the critical
temperature $T_{c}(\infty)$ at which some thermodynamic functions exhibit a
singularity is shifted to $T_{c} (L)$ critical temperature for a system
confined to the general geometry $L^{d-d^{\prime}}\!\!\times\!\!\infty^{d^{%
\prime}}$, when the system is infinite in $d^{\prime}$ dimensions and finite
in $(d-d^{\prime})$--dimensions. In the case when the number of infinite
dimensions is less than the lower critical dimension, there is no phase
transition in the system and the singularities of the thermodynamic
functions are altered. The critical temperature $T_{c}(\infty)$ in this case
is shifted to a pseudocritical temperature, corresponding to the center of
the rounding of the singularities of the thermodynamic functions, holding in
the thermodynamic limit.

In our quantum case, having in mind that we have considered the
low--temperature behavior of model~(\ref{model1}) in the context of the FSS
theory it is convenient to choose the quantum parameter $\lambda$ like a
critical instead of the temperature $t$ and to consider our system confined
to the geometry $L^{d-d^{\prime}}\!\!\times\!\!\infty^{d^{\prime}}\!\!\times%
\!\!L_{\tau}$. So the shifted critical quantum parameter $%
\lambda_{c}(t,L)\equiv\lambda_{tL}$ is obtained by setting $\phi=0$ in Eq.~(%
\ref{finite}). This gives 
\begin{equation}  \label{shift}
\frac{1}{\lambda_{tL}}-\frac{1}{\lambda_{c}} = \frac{\Gamma\left(\frac{d-1}{2%
}\right)} {4\pi^{(d+1)/2}} {\sum_{m,{\bbox l}(d-d^{\prime})}}^{\hskip%
-4mm\prime}\ \ \left((\lambda_{tL} m/t)^{2}+ (L|{\bbox l}|)^{2}%
\right)^{(1-d)/2}.
\end{equation}

The sum in the r.h.s. of Eq.~(\ref{shift}) is convergent for $d^{\prime}>2$,
however it can be expressed in terms of the Epstein zeta function (see Ref. %
\onlinecite{glasser80}) 
\begin{equation}  \label{epstein}
{\cal Z} \left| 
\begin{array}{c}
0 \\ 
0
\end{array}
\right|\left(L^{2}{\bbox l}^{2} +\left(\frac{\lambda}{t}\right)^{2}m^{2};d-1%
\right) = {\sum_{m,{\bbox l} (d-d^{\prime})}}^{\hskip-4mm\prime}\ \
\left(L^{2}{\bbox l}^{2}+\left(\frac{\lambda}{t}\right)^{2}m^{2} \right)^{%
\frac{1-d}{2}},
\end{equation}
which can be regarded as the generalized ($d-d^{\prime}+1$)-dimensional
analog of the Riemann zeta function $\zeta (\frac{d-1}{2})$. In the case
under consideration the Epstein zeta function has only a simple pole at $%
d^{\prime}= 2$ and may be analytically continued for $0 \leq d^{\prime}< 2$
to give a meaning to Eq.~(\ref{shift}) for $d^{\prime}<2$ as well. It is
hard to investigate the sum appearing in Eq.~(\ref{epstein}). The anisotropy
of the sum $L^{2}l_{1}^{2}+\cdots+L^{2}l_{d-d^{\prime}}^{2} +\left(\frac{%
\lambda}{t}\right)^{2}m^{2}$ is an additional problem. For these reasons we
will try to solve it asymptotically, considering different regimes of the
temperature, depending on that whether $L\ll\frac{\lambda_{tL}}{t}$ or $L\gg%
\frac{\lambda_{tL}}{t}$ which will be called, respectively, very
low--temperature regime and low--temperature regime. To do that it is more
convenient to write Eq.~(\ref{shift}) in the following form 
\begin{equation}  \label{start}
\frac{1}{\lambda_{tL}}-\frac{1}{\lambda_{c}} = \frac{1}{4\pi^{(d+1)/2}} {%
\sum_{m,{\bbox l}(d-d^{\prime})}}^{\hskip-4mm\prime}\ \ \int_{0}^{\infty} dx
x^{\frac{d-3}{2}}\exp\left\{-x\left[L^{2}{\bbox l}^{2}+\left(\frac{%
\lambda_{tL}}{t}\right)^{2}m^{2}\right]\right\}.
\end{equation}
This will be the starting equation for the two cases we will consider below.


\paragraph{Low--temperature regime $\frac{\lambda _{tL}}t\ll L$}

In this case we single out, from the sum in Eq.~(\ref{start}), the term with 
${\bbox l}=0$ to obtain 
\begin{eqnarray}
\frac{1}{\lambda_{tL}}-\frac{1}{\lambda_{c}}&=&\frac{1} {2\pi^{(d+1)/2}}%
\left(\frac{t}{\lambda_{tL}}\right)^{d-1} \Gamma\left(\frac{d-1}{2}%
\right)\zeta(d-1)  \nonumber \\
& &+\frac{1}{4\pi^{(d+1)/2}} {\sum_{{\bbox l}(d-d^{\prime})}}^{\hskip%
-2mm\prime} \sum_{m=-\infty}^{\infty} \int_{0}^{\infty} dx x^{\frac{d-3}{2}%
}\exp\left\{-x\left[L^{2} {\bbox l}^{2}+\left(\frac{\lambda_{tL}}{t}%
\right)^{2}m^{2}\right] \right\}.
\end{eqnarray}
After the application of the Jacobi identity (\ref{jacobi}) to the sum over $%
m$ and the calculation of the arising integrals in the resulting expression
we obtain the final result 
\begin{eqnarray}  \label{shift1}
\frac{1}{\lambda_{tL}}-\frac{1}{\lambda_{c}} &=&\frac{1}{2 \pi^{(d+1)/2}}%
\left(\frac{t}{\lambda_{tL}}\right)^{d-1} \Gamma\left(\frac{d-1}{2}%
\right)\zeta(d-1) +\frac{t}{\lambda_{tL}}\frac{L^{2-d}}{4\pi^{d/2}}
\Gamma\left(\frac{d}{2}-1\right) {\sum_{{\bbox l}(d-d^{\prime})}}^{\hskip%
-2mm\prime} |{\bbox l}|^{2-d}  \nonumber \\
& &+\left(\frac{t}{\lambda_{tL}}\right)^{d/2}\frac{L^{1-d/2}}{\pi} {\sum_{{%
\bbox l}(d-d^{\prime})}}^{\hskip-2mm\prime}\ \ \sum_{m=1}^{\infty} \left(%
\frac{m}{|{\bbox l}|}\right)^{d/2-1}K_{\frac{d}{2}-1} \left(2\pi\frac{t}{%
\lambda_{tL}}Lm|{\bbox l}| \right).
\end{eqnarray}

The first term of the r.h.s of Eq.~(\ref{shift1}) is the shift of the critical
quantum parameter (see Eq.~(\ref{bulk10})) due to the presence of the
quantum effects in the system. The second term is a correction resulting
from the finite sizes. It is just the shift due the finite--size effects in
the $d$--dimensional spherical model~\cite{chamati96} multiplied by the
temperature scaled to the quantum parameter. Here the $(d-d^{\prime})$--fold
sum may be continued analytically beyond its domain of convergence with
respect to ``$d$'' and ``$d^{\prime}$'' (which is $2<d^{\prime}<d$). The last
term is exponentially small in the considered limit i.e. $\frac{\lambda_{tL}%
}{t}\ll L$.

In the borderline case $d=2$, one can see that in the first and second
terms in the r.h.s of Eq.~(\ref{shift1}) singularities take place. In Appendix~%
\ref{app3} we can show that they cancel and Eq.~(\ref{shift1}) then yields 
\begin{equation}  \label{gammae}
\frac{1}{\lambda_{tL}}-\frac{1}{\lambda_{c}}=\frac{t}{2\pi\lambda_{tL}}
\left\{B_{0}+\gamma_{E}^{}+\ln\frac{tL}{2\lambda_{tL}}+2 {\sum_{{\bbox l}
(2-d^{\prime})}}^{\hskip-2mm\prime}\sum_{m=1}^{\infty} K_{0}\left(2\pi\frac{t%
}{\lambda_{tL}}Lm|{\bbox l}| \right)\right\},
\end{equation}
where $\gamma_{E}^{}=.577...$ is the Euler constant.

In the particular case of strip geometry $d^{\prime}=1$, the shift is given
by (see Appendix~\ref{app3}) 
\begin{equation}  \label{lt<1}
\frac{1}{\lambda_{tL}}-\frac{1}{\lambda_{c}}=\frac{t}{2\pi\lambda_{tL}}
\left\{\gamma_{E}^{}+\ln\frac{tL}{4\pi\lambda_{tL}}+4\sum_{l=1}
^{\infty}\sum_{m=1}^{\infty}K_{0}\left(2\pi\frac{t}{\lambda_{tL}}Lml
\right)\right\}.
\end{equation}

For the fully finite geometry case ($d^{\prime}=0$) we obtain for the shift
(see Appendix~\ref{app3}) 
\begin{equation}  \label{tl<1}
\frac{1}{\lambda_{tL}}-\frac{1}{\lambda_{c}}=\frac{t}{2\pi \lambda_{tL}}%
\left\{\gamma_{E}^{}+\ln\frac{tL}{\lambda_{tL}}- \ln\frac{\left[\Gamma\left(%
\frac{1}{4}\right)\right]^{2}}{\sqrt{\pi}}+ 2 {\sum_{l_{1},l_{2}}}%
^{\prime}\sum_{m=1}^{\infty} K_{0}\left(2\pi\frac{t}{\lambda_{tL}}Lm\sqrt{%
l_{1}^{2}+l_{2}^{2}} \right)\right\}.
\end{equation}
Let us note that the last term in the r.h.s of Eqs.~(\ref{lt<1}) and~(\ref
{tl<1}) gives exponentially small corrections and in the considered limit
they can be omitted.

\paragraph{Very low--temperature regime $\frac \lambda t\gg L$}

To get an appropriate expression now, we single out the term corresponding
to $m=0$ from the sum in Eq.~(\ref{start}). This leads to 
\begin{eqnarray}
\frac{1}{\lambda_{tL}}-\frac{1}{\lambda_{c}} &=& \frac{L^{1-d}}{%
4\pi^{(d+1)/2}} \Gamma\left(\frac{d-1}{2}\right) {\sum_{{\bbox l}%
(d-d^{\prime})}}^{\hskip-2mm\prime}\ \ |{\bbox l}|^{1-d}  \nonumber \\
& &+\frac{1}{2\pi^{(d+1)/2}} \sum_{{\bbox l}(d-d^{\prime})}\sum_{m=1}^{%
\infty} \int_{0}^{\infty}dx\ x^{\frac{d-3}{2}}\exp\left\{-x\left[L^{2} {%
\bbox l}^{2}+\left(\frac{\lambda_{tL}}{t}\right)^{2}m^{2} \right]\right\}.
\end{eqnarray}
The next step consists in the application of the Jacobi identity to the $%
(d-d^{\prime})$--dimensional sum and the calculation of the arising
integrals to obtain 
\begin{eqnarray}  \label{shift2}
\frac{1}{\lambda_{tL}}-\frac{1}{\lambda_{c}} &=& \frac{L^{1-d}}{%
4\pi^{(d+1)/2}} \Gamma\left(\frac{d-1}{2}\right) {\sum_{{\bbox l}%
(d-d^{\prime})}}^{\hskip-2mm\prime}\ \ |{\bbox l}|^{1-d} +\frac{%
L^{d^{\prime}-d}}{2\pi^{(d^{\prime}+1)/2}}\Gamma\left(\frac{d^{\prime}-1}{2}%
\right) \left(\frac{t}{\lambda_{tL}}\right)^{d^{\prime}-1}\zeta(d^{\prime}-1)
\nonumber \\
& &+\frac{L^{1/2+d^{\prime}/2-d}}{\pi} \left(\frac{t}{\lambda_{tL}} \right)^{%
\frac{d^{\prime}-1}{2}}{\sum_{{\bbox l}(d-d^{\prime})}}^{\hskip-2mm\prime}\
\ \sum_{m=1}^{\infty}\left(\frac{|{\bbox l}|}{m}\right)^{\frac{d^{\prime}-1}{%
2}}K_{\frac{d^{\prime}-1}{2}} \left(2\pi\frac{\lambda_{tL}}{tL}m|{\bbox l}|
\right).
\end{eqnarray}
Here, in the r.h.s, the first term is the expression of the shift of the
critical quantum parameter, at zero temperature,~\cite{chamati94} due to the
finite sizes of the system. This is equivalent to the shift of a ($d+1$%
)--dimensional spherical model confined to the geometry $L^{d+1-d^{\prime}}%
\!\!\times\!\!\infty^{d^{\prime}}$. The second term gives corrections due
to the quantum effects. This is the shift of critical quantum parameter of a 
$d^{\prime}$--dimensional infinite system multiplied by the volume of a $%
(d-d^{\prime})$--dimensional hypercube. The third term is exponentially
small in the limit of very low temperatures. The singularity for $%
d^{\prime}=1$ of the r.h.s of Eq.~(\ref{shift2}) is fictitious. So for $%
d^{\prime}=1$ Eq.~(\ref{shift2}) yields (see Appendix~\ref{app3}) 
\begin{equation}  \label{d'=1}
\frac{1}{\lambda_{tL}}-\frac{1}{\lambda_{c}}=\frac{L^{1-d}}{2\pi} \left\{\ln%
\frac{\lambda_{tL}}{2\sqrt{\pi}tL}+\frac{\gamma_{E}^{}}{2}+ \tilde{C_{0}}+2{%
\sum_{{\bbox l}(d-1)}}^{\hskip-2mm\prime} \sum_{m=1}^{\infty}K_{0}\left(2\pi%
\frac{\lambda_{tL}}{tL}m| {\bbox l}| \right)\right\},
\end{equation}
where 
$$
\tilde{C_{0}}=\left[2\pi^{(d-1)/2}\right]^{-1} \Gamma\left(\frac{d-1}{2}%
\right) C_{0}, 
$$
In the particular case $d^{\prime}=1$ and $d=2$ the constant $C_{0}$ is
given by $C_{0}=\gamma_{E}^{}-\ln4\pi$ and 
\begin{equation}  \label{tl>1}
\frac{1}{\lambda_{tL}}-\frac{1}{\lambda_{c}}=\frac{1}{2\pi L}\left
\{\gamma_{E}^{}+\ln\frac{\lambda_{tL}}{4\pi tL}+ 4\sum_{l=1}^{\infty}
\sum_{m=1}^{\infty} K_{0}\left(2\pi\frac{\lambda_{tL}}{tL}ml\right)\right\}.
\end{equation}

Comparing between Eqs.~(\ref{lt<1}) and~(\ref{tl>1}) one can see the crucial
role (in symmetric form) of $L$ or $\lambda_{tL}/t$ in the low--temperature
regime and the very low--temperature one, respectively.

In another particular case of a two--dimensional bloc geometry $%
d^{\prime}=0$ and $d=2$, using the Hardy formula (see Appendix~\ref{app3})
one gets 
\begin{equation}  \label{dp=0}
\frac{1}{\lambda_{tL}}-\frac{1}{\lambda_{c}}=\frac{L^{-1}}{\pi} \zeta\left(%
\frac{1}{2}\right)\beta\left(\frac{1}{2}\right) -\frac{\lambda_{tL}}{tL^{2}}%
\zeta(-1)+\frac{L^{-1}}{2\pi} \sum_{m=1}^{\infty}{\sum_{l_{1},l_{2}}}%
^{\prime}\frac{ \exp\left[\frac{-2\pi\lambda_{tL}}{tL}%
m\left(l_{1}^{2}+l_{2}^{2} \right)^{1/2}\right]}{\sqrt{l_{1}^{2}+l_{2}^{2}}}.
\end{equation}

Instead of the previous case of low--temperature regime here the lower
quantum critical dimension $d^{\prime}=1$ is responsible for the logarithmic
dependence in Eq.~(\ref{d'=1}). This is the reason for the significant
difference between Eqs.~(\ref{tl>1}) and~(\ref{dp=0}).

The above obtained equations for $\lambda_{tL}$ will be exploited later for
the study of the two dimensional case.

\subsubsection{Zero--field susceptibility}\label{zerofs}
It is possible to transform Eq.~(\ref{finite}) in the
following equivalent forms 
\begin{mathletters}
\label{fssl}
\begin{equation}
L^{d-1}\delta\lambda+\left(\frac{h L^{(d+3)/2}}{L^{2} \phi} \right)^{2}=%
\frac{(L\phi^{1/2})^{d-1}}{(4\pi)^{(d+1)/2}} \left[\left|\Gamma\left(\frac{%
1-d}{2}\right)\right|- 2{\cal K}_{\frac{\lambda}{tL}}\left.\left(\frac{d-1}{2%
}\right| d-d^{\prime}+1,\frac{L\phi^{1/2}}{2}\right)\right],
\end{equation}
where 
\begin{equation}  \label{kat}
{\cal K}_{a} (\nu|p,y) = {\sum_{m,{\bbox l}(p-1)}}^{\hskip-3.5mm\prime} 
\frac{K_{\nu}\left(2y\sqrt{{\bbox l}^{2}+a^{2}m^{2}}\right)} {\left(y\sqrt{{%
\bbox l}^{2}+a^{2}m^{2}}\right)^{\nu}}, \ \ y>0, \ \ {\bbox l}^2 =
l_{1}^{2}+l_{2}^{2}+\cdots+l_{p-1}^{2}.
\end{equation}
or 
\end{mathletters}
\begin{mathletters}\label{fsst}
\begin{eqnarray}
\left(\frac{t}{\lambda}\right)^{1-d}\delta\lambda +\left[\frac{h}{\phi}\left(%
\frac{t}{\lambda}\right)^{2} \left(\frac{\lambda}{t}\right)^{(d+3)/2}%
\right]^{2}&=& \frac{(\lambda\phi^{1/2}/t)^{d-1}}{(4\pi)^{(d+1)/2}}
\left[\left|\Gamma\left(\frac{1-d}{2}\right)\right|\right.  \nonumber \\
& &\left.-2\tilde{{\cal K}}_{\frac{tL}{\lambda}} \left.\left(\frac{d-1}{2}%
\right|d-d^{\prime}+1,\frac{\lambda\phi^{1/2}}{2t} \right)\right],
\end{eqnarray}
where 
\begin{equation}  \label{katt1}
\tilde{{\cal K}}_{a} (\nu|p,y) = {\cal K}_{\frac{1}{a}} (\nu|p;ay) ={\sum_{m,%
{\bbox l}(p-1)}}^{\hskip-3.5mm\prime}\ \ \frac{K_{\nu}\left(2y\sqrt{a^{2}{%
\bbox l}^{2}+m^{2}}\right)} {\left(y\sqrt{a^{2}{\bbox l}^{2}+m^{2}}%
\right)^{\nu}}, \ \ y>0.
\end{equation}
The functions ${\cal K}_{a} (\nu|p;y)$ and $\tilde{{\cal K}}_{a} (\nu|p;y)$
are anisotropic generalizations of the ${\cal K}$--function introduced in
Ref.~\onlinecite{singh85}.

Eqs.~(\ref{fssl}) and~(\ref{fsst}) show that the correlation length $%
\xi=\phi^{-1/2}$ will scale like 
\end{mathletters}
\begin{mathletters}
\begin{equation}
\xi=L f_{\xi}^{L} \left\{\delta\lambda L^{1/\nu},\frac{tL}{\lambda},
hL^{\Delta/\nu}\right\},
\end{equation}
or like 
\begin{equation}
\xi=\frac{\lambda}{t}f_{\xi}^{t} \left\{\delta\lambda\left(\frac{t} {\lambda}%
\right)^{-1/\nu},\frac{tL}{\lambda},h\left(\frac{t}{\lambda}
\right)^{-\Delta/\nu}\right\},
\end{equation}
which suggests also that there will be some kind of interplay (competition)
between the finite--size and the quantum effects.

Hereafter we will try to find the behavior of the susceptibility $%
\chi=\phi^{-1}$ as a function of the temperature $t$ and the size $L$ of the
system. For simplicity, in the remainder of this section, we will
investigate the free field case ($h=0$).

{\it 1.} For $\frac{\lambda}{t}\phi^{1/2}\ll1$, after using the asymptotic
form of the function defined in~(\ref{kat}) (see Appendix~\ref{app2})
\end{mathletters}
\begin{equation}
{\cal K}_{a} (\nu|p,y)\approx\frac{\pi^{p/2}}{2a} \Gamma\left( \frac{p}{2}%
-\nu\right)y^{-p}+\frac{\pi^{2\nu-p/2}}{2a} y^{-2\nu} C_{a}(p|\nu)-\frac{1}{2%
} \Gamma(-\nu),
\end{equation}
Eq.~(\ref{finite}) reads ($d^{\prime}\neq2$, $1<d<3$) 
\begin{equation}  \label{fssa}
\delta\lambda+\frac{t}{\lambda}\frac{L^{d^{\prime}-d}}{(4\pi)^{d^{\prime}/2}}
\Gamma\left(1-\frac{d^{\prime}}{2}\right)\phi^{(d^{\prime}-2)/2}+\frac{1} {%
4\pi^{(d+1)/2}}\Gamma\left(\frac{d-1}{2}\right) {\sum_{m,{\bbox l}%
(d-d^{\prime})}}^{\hskip-4mm\prime} \left[\left(\frac{\lambda}{t}%
m\right)^{2}+\left(L{\bbox l}\right)^{2} \right]^{\frac{1-d}{2}}=0.
\end{equation}

Now we will examine Eq.~(\ref{fssa}) in different regimes of $t$ and $L$ and
for different geometries of the lattice:

{\it 1.a.} $\frac{\lambda}{t}\phi^{1/2}\ll1$ and $\frac{tL}{\lambda}\gg1$:
In this case Eq.~(\ref{fssa}) transforms into (up to exponentially small
corrections c.f. with Eq.~(\ref{shift1})) 
\begin{eqnarray}  \label{lta}
0&=&\delta\lambda+\frac{t}{\lambda}\frac{L^{d^{\prime}-d}}{%
(4\pi)^{d^{\prime}/2}} \Gamma\left(1-\frac{d^{\prime}}{2}\right)\phi^{(d^{%
\prime}-2)/2}+ \frac{1}{2\pi^{(d+1)/2}}\left(\frac{t}{\lambda}\right)^{d-1}
\Gamma\left(\frac{d-1}{2}\right)\zeta(d-1)  \nonumber \\
& &+\frac{t}{\lambda}\frac{L^{2-d}}{4\pi^{d/2}}\Gamma\left(\frac{d}{2}
-1\right){\sum_{{\bbox l}(d-d^{\prime})}}^{\hskip-2mm\prime} \ \ |{\bbox l}%
|^{2-d}.
\end{eqnarray}
This equation has different types of solutions depending on whether the
dimensionality $d$ is above or below the classical critical dimension 2.

At $\lambda=\lambda_{c}$ and when $d^{\prime}<2<d<3$ (i.e. when there is no
phase transition in the system) we obtain for the zero--field susceptibility 
\begin{equation}  \label{ltra}
\chi=\left(\frac{t}{\lambda_{c}}\right)^{-2} \left(\frac{tL}{\lambda_{c}}%
\right)^{2(d-d^{\prime})/(2-d^{\prime})} \left[\frac{2^{d^{\prime}-1}}{%
\pi^{(d-d^{\prime}+1)/2}}\frac{\Gamma\left( \frac{d-1}{2}\right)}{%
\Gamma\left(1-\frac{d^{\prime}}{2}\right)} \zeta(d-1)\right]^{\frac{2}{%
2-d^{\prime}}}.
\end{equation}
However, for $1<d<2$, Eq.~(\ref{lta}) has no solution at $\lambda=\lambda_{c}$
obeying the initial condition $\frac{\lambda_{c}}{t}\phi^{1/2}\ll1$.

Eq.~(\ref{ltra}) generalizes the bulk result~(\ref{bulk8}) for $d$ close to
the upper quantum critical dimension i.e. $d=3$.

At the shifted critical quantum parameter $\lambda_{c}(t)$ given by Eq.~(\ref
{bulk10}) we get 
\begin{equation}  \label{zin}
\chi=L^{2} \left[\frac{2^{d^{\prime}-2}}{\pi^{(d-d^{\prime})/2}} \frac{%
\Gamma\left(\frac{d}{2}-1\right)}{\Gamma\left(1-\frac{d^{\prime}}{2} \right)}%
{\sum_{{\bbox l}(d-d^{\prime})}}^{\hskip-2mm\prime}\ \ |{\bbox l}|^{2-d}
\right]^{\frac{2}{2-d^{\prime}}}.
\end{equation}
However this solution is valid only for $3>d>2>d^{\prime}$ i.e. here again
there is no phase transition in the system.

{\it 1.b.} $\frac{\lambda}{t}\phi^{1/2}\ll1$ and $\frac{tL}{\lambda}\ll1$:
In this case, Eq.~(\ref{fssa}) gives (up to exponentially small corrections
c.f. with Eq.~(\ref{shift2})) 
\begin{eqnarray}  \label{vlta}
0&=&\delta\lambda+\frac{t}{\lambda}\frac{L^{d^{\prime}-d}}{%
(4\pi)^{d^{\prime}/2}} \Gamma\left(1-\frac{d^{\prime}}{2}\right)\phi^{(d^{%
\prime}-2)/2}+ \frac{L^{1-d}}{4\pi^{(d+1)/2}} \Gamma\left(\frac{d-1}{2}%
\right) {\sum_{{\bbox l}(d-d^{\prime})}}^{\hskip-2mm\prime}\ \ |{\bbox l}%
|^{1-d}  \nonumber \\
& &+\frac{L^{d^{\prime}-d}}{2\pi^{(d^{\prime}+1)/2}}\Gamma\left(\frac{%
d^{\prime}-1}{2}\right) \left(\frac{t}{\lambda}\right)^{d^{\prime}-1}%
\zeta(d^{\prime}-1).
\end{eqnarray}
Here we find that the solutions of Eq.~(\ref{vlta}) depend upon whether
the dimensionality $d^{\prime}<1$ or $d^{\prime}>1$.

At $\lambda=\lambda_{c}$ and for $1<d^{\prime}<2$, Eq.~(\ref{vlta}) has 
\begin{equation}  \label{vltra}
\chi=L^{2}\left(\frac{\lambda_{c}}{tL}\right)^{2/(2-d^{\prime})} \left[\frac{%
2^{d^{\prime}-2}}{\pi^{(d-d^{\prime}+1)/2}} \frac{\Gamma\left(\frac{d-1}{2}%
\right)}{\Gamma\left(1-\frac{d^{\prime}}{2} \right)}{\sum_{{\bbox l}%
(d-d^{\prime})}}^{\hskip-2mm\prime}\ \ |{\bbox l}|^{1-d} \right]^{\frac{2}{%
2-d^{\prime}}}
\end{equation}
as a solution. For $0\leq d^{\prime}<1$, however, it has no solution obeying
the initially imposed restriction $\frac{\lambda_{c}}{t}\phi^{1/2}\ll1$.

At the shifted critical quantum parameter $\lambda_{c}(L)$ given by~\cite
{chamati94} 
$$
\frac{1}{\lambda}-\frac{1}{\lambda_{c}(L)}=\frac{L^{1-d}} {4\pi^{(d+1)/2}}%
\Gamma\left(\frac{d-1}{2}\right) {\sum_{{\bbox l}(d-d^{\prime})}}^{\hskip%
-2mm\prime} |{\bbox l}|^{1-d}, 
$$
Eq.~(\ref{vlta}) has a solution obeying the initial condition $\frac{%
\lambda_{c}}{t}\phi^{1/2}\ll1$ only for $d^{\prime}=1+\varepsilon$ and in
this case the susceptibility behaves like 
\begin{equation}  \label{epst}
\chi=\frac{1}{(\pi\varepsilon)^{2}}\frac{\lambda_{c}^{2}}{t^{2}}
\left[1-\varepsilon\left(\gamma_{E}^{}+\ln\frac{\varepsilon} {2}%
\right)\right]^{2}.
\end{equation}

{\it 2.} For $L\phi^{1/2}\ll1$ from Eqs.~(\ref{fssl}) and Eq.~(\ref{lastap})
we get once again Eq.~(\ref{lta}). In spite of the fact that we have the same
equation as in the case $\frac{\lambda}{t}\phi^{1/2}\ll1$, the expected
solutions for the susceptibility may be different because of the new imposed
condition. Here also we will consider the two limiting cases of
low--temperature and very low--temperature regimes.

{\it 2.a.} $L\phi^{1/2}\ll1$ and $\frac{tL}{\lambda}\gg1$: In this case Eq.~(%
\ref{fssa}) again is transformed into Eq.~(\ref{lta}) and we obtain at $%
\lambda=\lambda_{c}$ the solution given by Eq.~(\ref{ltra}), which is valid
only for $d^{\prime}<2<d<3$, i.e. we have the same solution as in the
previous case i.e. $\frac{\lambda}{t}\phi^{1/2}\ll1$.

At $\lambda=\lambda_{c}(t)$, we formally obtain Eq.~(\ref{zin}) which,
however, may be considered like a solution only in the neighborhood of the
lower classical critical dimension $d=2$. For the cylindric geometry ($%
d^{\prime}=1$ and $d=2+\varepsilon$) we get 
\begin{equation}  \label{cy}
\chi=\frac{L^{2}}{(\pi\varepsilon)^{2}}\left[1-\frac{\varepsilon}{2}
(\gamma_{E}^{}-\ln4\pi)\right]^{2}.
\end{equation}
This result is contained in Eq. (30.109) of Ref.~\onlinecite{Zin89} in the
large $n$--limit case for the NLQ$\sigma$M.

In the case of slab geometry $d-d^{\prime}=1$ ($d=2+\varepsilon,
d^{\prime}=1+\varepsilon$) instead of~(\ref{cy}) we obtain 
\begin{equation}  \label{cy1}
\chi=\frac{L^{2}}{(\pi\varepsilon)^{2}}\left[1-\varepsilon(
\gamma_{E}^{}-\ln2)-\varepsilon\ln\varepsilon\right]^{2}.
\end{equation}

In the case of a bloc geometry ($d=2+\varepsilon$ and $d^{\prime}=0$) we
find the following behavior for the susceptibility 
\begin{equation}
\chi=\frac{L^{2}}{2\pi\varepsilon}\left[1-\frac{\varepsilon}{4}\left(
\gamma_{E}^{}-\ln\frac{\left[\Gamma\left(\frac{1}{4}\right)\right]^{4}} {%
4\pi^{2}}\right)\right]^{2}.
\end{equation}

For the case of ``quasi--bloc geometry'' ($d=2+\varepsilon$ and $%
d^{\prime}=\varepsilon$) we get 
\begin{equation}  \label{epsb}
\chi=\frac{L^{2}}{2\pi\varepsilon}\left[1-\frac{\varepsilon}{4}\left(
2\gamma_{E}^{}+\ln\frac{\pi\varepsilon}{2}-2\ln\frac{\left[\Gamma\left( 
\frac{1}{4}\right)\right]^{2}}{2\sqrt{\pi}}\right)\right]^{2}.
\end{equation}
Here the appearance of $\varepsilon$ in the denominator in formulas~(\ref
{epst})--(\ref{epsb}) signalizes that the scaling in its simple form will
fail at $\varepsilon=0$.

{\it 2.b.} $L\phi^{1/2}\ll1$ and $\frac{tL}{\lambda}\ll1$: Here we find that
Eq.~(\ref{vlta}) is valid, and it has Eq.~(\ref{vltra}) as a solution at $%
\lambda=\lambda_{c}$ and for $0\leq d^{\prime}<1$. For $1<d^{\prime}<2$ the
susceptibility is given by 
\begin{equation}  \label{vltrb}
\chi=\left(\frac{\lambda_{c}}{2t}\right)^{2} \left[\frac{2}{\pi^{1/2}} \frac{%
\Gamma\left(\frac{d^{\prime}-1}{2}\right)}{\Gamma\left(1-\frac{d^{\prime}}{2}
\right)}\zeta(d^{\prime}-1)\right]^{\frac{2}{2-d^{\prime}}}.
\end{equation}

At the shifted critical point $\lambda_{c}(L)$, for the susceptibility we
obtain Eq.~(\ref{vltrb}) under the restriction $2>d^{\prime}>1$, which
guarantees the positiveness of the quantity under brackets.

When $\lambda<\lambda_{c}$ for $1<d<3$ and $d^{\prime}<2$, i.e. when there
is no phase transition in the system, we obtain 
\begin{equation}  \label{bussa}
\chi=\left[\frac{(4\pi)^{d^{\prime}/2}}{\Gamma\left(1-\frac{d^{\prime}}{2}%
\right)} \left(1-\frac{\lambda}{\lambda_{c}}\right)\right] ^{\frac{2}{%
2-d^{\prime}}} t^{-2/(2-d^{\prime})} L^{2(d-d^{\prime})/(2-d^{\prime})}.
\end{equation}

If $d^{\prime}>2$ there is a phase transition in the system at the shifted
value of the critical quantum parameter $\lambda_{tL}$ (the shift in this
case is due to the quantum and finite--size effects) and Eq.~(\ref{fssa})
transforms to 
\begin{equation}
1-\frac{\lambda}{\lambda_{tL}}=t\frac{L^{d^{\prime}-d}}{(4\pi)^{d/2}}
\Gamma\left(1-\frac{d^{\prime}}{2}\right)\phi^{(d^{\prime}-2)/2},
\end{equation}
which has the following solutions 
\begin{equation}  \label{bussb}
\chi=\left\{ 
\begin{array}{ll}
\left[\frac{(4\pi)^{d^{\prime}/2}}{\Gamma\left(1-\frac{d^{\prime}}{2}\right)}
\left(1-\frac{\lambda}{\lambda_{tL}}\right)\right] ^{\frac{2}{2-d^{\prime}}}
t^{-2/(2-d^{\prime})} L^{2(d-d^{\prime})/(2-d^{\prime})}, & 
\lambda>\lambda_{tL} \\ 
\infty , & \lambda\leq\lambda_{tL}
\end{array}
\right.
\end{equation}
Let us notice that Eqs.~(\ref{bussa}) and~(\ref{bussb}) are the finite--size
forms, for the susceptibility, of Eqs.~(\ref{suscepb}) and~(\ref{bulk11}),
respectively, found for the bulk system.


\subsubsection{Two--dimensional case}\label{tdc}
The two--dimensional case needs special treatment because of its
physical reasonability and the increasing interest in the context of the
quantum critical phenomena.~\cite
{chakravarty89,chubukov94,sachdev94,Rosenstein90,neuberger89,fisher89,hasenfratz93,azaria93,Castro93,fujii95}
From Eq.~(\ref{finite}) for $d=2$ and in the absence of a magnetic field $%
h=0 $ we get 
\begin{equation}  \label{2dcase}
\delta\lambda=\frac{\phi^{1/2}}{4\pi}-\frac{1}{4\pi} {\sum_{m,{\bbox l}%
(2-d^{\prime})}}^{\hskip-4mm\prime}\ \ \frac{\exp\left[-\phi^{1/2}\left(%
\frac{\lambda^{2}}{t^{2}}m^{2} + L^{2}{\bbox l}^{2}\right)^{1/2}\right]}{%
\left(\frac{\lambda^{2}} {t^{2}}m^{2} + L^{2}{\bbox l}^{2}\right)^{1/2}}.
\end{equation}
Introducing the scaling functions $Y_{t}^{d^{\prime}}=\frac{\lambda}{t}%
\phi^{1/2}$ and $Y_{L}^{d^{\prime}}=L\phi^{1/2}$, where the superscript $%
d^{\prime}$ denotes the number of infinite dimensions in the system, and the
scaling variable $a=\frac{tL}{\lambda}$ it is easy to write Eq.~(\ref{2dcase}%
) in the scaling forms given in Eqs.~(\ref{fssl}) and (\ref{fsst}). The
solutions of the obtained scaling equations will depend on the number of the
infinite dimensions in the system. Here we will consider the two most
important particular cases: strip geometry $d^{\prime}=1$ and bloc geometry $%
d^{\prime}=0$. Our analysis will be confined to the study of the behavior
of the scaling functions at the critical value of the quantum parameter $%
\lambda_{c}$, and at the shifted critical quantum parameter $\lambda_{tL}$
(see Section~\ref{scqp}). It is difficult to solve Eq~(\ref{2dcase}) by
using an analytic approach that is for what we will give a numerical
treatment of the problem. It is, however, possible to consider the two
limits: $a\gg1$ i.e. the low--temperature regime and $a\ll1$ i.e. the very
low--temperature regime.

\begin{figure}[tbp]
\vspace{5cm}
\epsfxsize=4.5in
\centerline{\epsffile{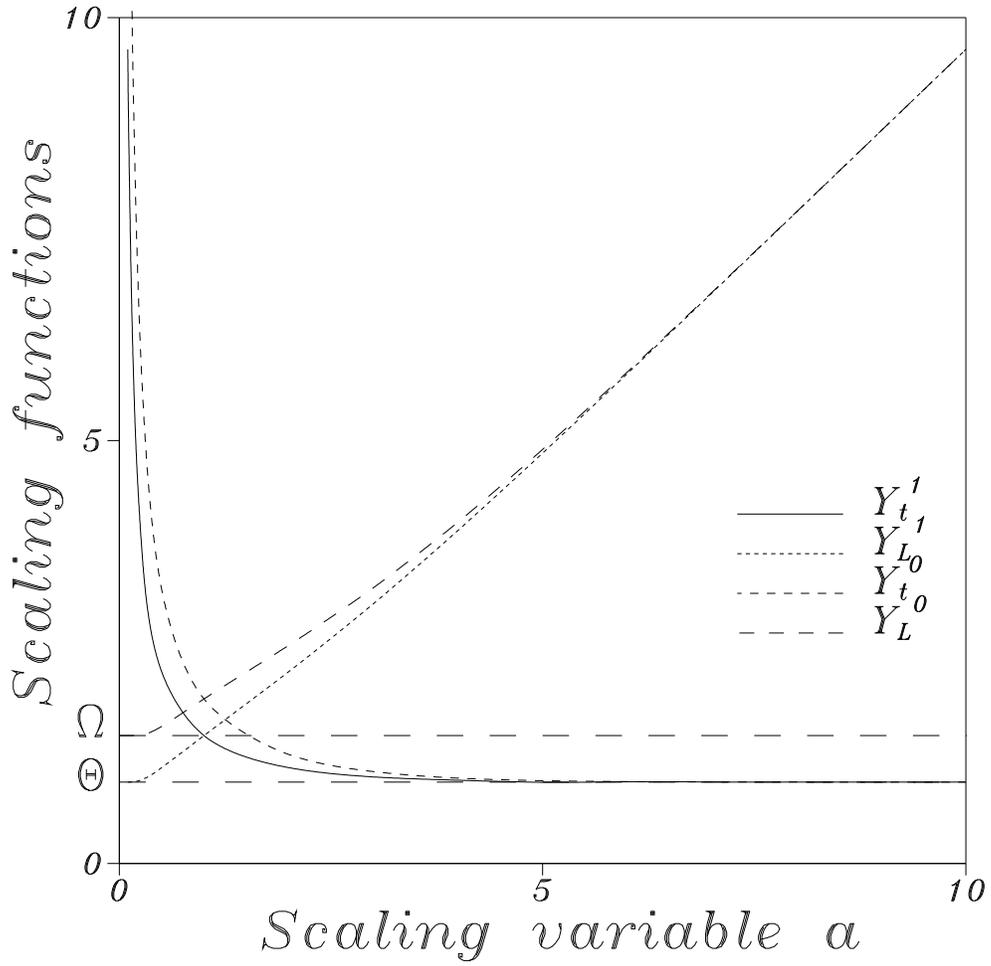}}
\vspace{0.1in}
\caption{The effects of the finite--size geometry on the bulk behavior of $
\phi^{1/2}$ for the two dimensional case at $\lambda=\lambda_{c}$. The
superscript $d^{\prime}$ in $Y^{d^{\prime}}_{L}=L\phi^{1/2}$ and $
Y^{d^{\prime}}_{t}=\frac{\lambda_{c}}{t}\phi^{1/2}$ indicates the number of
infinite dimensions in the system. The scaling variable $a=\frac{tL}{
\lambda_{c}}$. The universal numbers are $\Theta=0.962424...$ (see
Eq.~(\protect\ref{theta})) and $\Omega=1.511955...$}
\label{fig2}
\end{figure}

{\it Strip geometry ($d^{\prime}=1$)}: In this case in the r.h.s of Eq.~(\ref
{2dcase}) we have a two--fold sum which permits a numerical analysis of the
geometry under consideration. FIG.~\ref{fig2} graphs the variation of the
scaling functions $Y_{t}^{1}$ and $Y_{L}^{1}$ against the variable $a$ at $%
\lambda=\lambda_{c}$. This shows that for a comparatively small value of the
scaling variable $a\sim5$ the finite--size behavior (see the curve of the
function $Y_{t}^{1}(a)$) merges in the low temperature bulk one, while the
behavior of $Y_{L}^{1}(a)$ shows that for relatively not very
low--temperatures ($a\sim\frac{1}{5}$, $L$-fixed) the system simulates the
behavior of a three--dimensional classical spherical model with one finite
dimensions. The mathematical reasons for this are the exponentially small
values of the corrections, as we will show below.

{\it Bloc geometry ($d^{\prime}=0$)}: In this case the three--fold sum in
the r.h.s of Eq.~(\ref{2dcase}) is not an obstacle to analyze it numerically.
For $\lambda=\lambda_{c}$ the behavior of the scaling functions $%
Y_{L}^{0}(a)$ and $Y_{t}^{0}(a)$ is presented in FIG.~\ref{fig2}. They have
the same qualitative behavior as in the strip geometry the only difference
is the appearance of a new universal number for $t=0$, i.e. $\Omega$,
instead of the constant $\Theta$ as a consequence of the asymmetry of the
sum in the low--temperature and the very low--temperature regimes.

Analytically for arbitrary values of the number of infinite dimensions $%
d^{\prime}$ we can treat first the problem in the low--temperature regime ($%
a\gg1$). In this limit Eq.~(\ref{2dcase}) can be transformed into (up to
small corrections ${\cal O}(e^{-2\pi a})$) 
\begin{equation}  \label{2da>1}
\delta\lambda=\frac{1}{2\pi\lambda}\ln2\sinh \frac{\lambda}{2t}\phi^{1/2}-%
\frac{1}{2\pi\lambda}{\sum_{{\bbox l}(2-d^{\prime})} }^{\hskip%
-2mm\prime}K_{0}\left(L\phi^{1/2}|{\bbox l}|\right).
\end{equation}
For $\lambda=\lambda_{c}$ Eq.~(\ref{2da>1}) has the solution 
\begin{equation}
\chi^{-1/2}\approx\frac{t}{\lambda_{c}}\Theta + (2-d^{\prime})\sqrt{\frac{%
2\pi}{5\Theta}}\left(\frac{t}{L\lambda_{c}} \right)^{1/2}\exp\left(-\frac{tL%
}{\lambda_{c}}\Theta\right)
\end{equation}
i.e. the finite--size corrections to the bulk behavior are exponentially
small.

In the very low--temperature regime ($a\ll1$), Eq.~(\ref{2dcase}) reads (up
to ${\cal O}(e^{-2\pi/a})$) 
\begin{equation}  \label{2da<1}
\delta\lambda=\frac{\phi^{1/2}}{4\pi}-\frac{L^{-1}}{4\pi}{\sum_{{\bbox l}
(2-d^{\prime})}}^{\hskip-2mm\prime} \frac{\exp\left(-L\phi^{1/2}|{\bbox l}%
|\right)}{|{\bbox l}|}+ \frac{L^{d^{\prime}-2}}{\pi^{(d^{\prime}+1)/2}}%
\left(2\frac{\lambda}{t}\right) ^{\frac{1-d^{\prime}}{2}}\sum_{m=1}^{%
\infty}K_{\frac{d^{\prime}-1}{2}} \left(\frac{\lambda}{t}\phi^{1/2}\right),
\end{equation}
which has the solutions: 
\begin{equation}
\chi^{-1/2}\approx\frac{1}{L}\Theta + \sqrt{\frac{2\pi}{5\Theta}}\left(\frac{%
L\lambda_{c}}{t}\right)^{1/2} \exp\left(-\frac{\lambda_{c}}{tL}\Theta\right)
\end{equation}
for $d^{\prime}=1$, and 
\begin{equation}  \label{Omega}
\chi^{-1/2}\approx\frac{1}{L}\Omega+\frac{1}{L}\left\{\frac{1}{2\Omega} +%
\frac{\Omega}{2}{\sum_{{\bbox l}(2)}}^{\prime} \left(\Omega^{2}+ 4\pi^{2}{%
\bbox l}^{2}\right)^{-3/2}\right\}^{-1}\exp\left(-\Omega\frac {\lambda_{c}}{%
tL}\right)
\end{equation}
for $d^{\prime}=0$.

\begin{figure}[tbp]
\vspace{5cm}
\epsfxsize=4.5in
\centerline{\epsffile{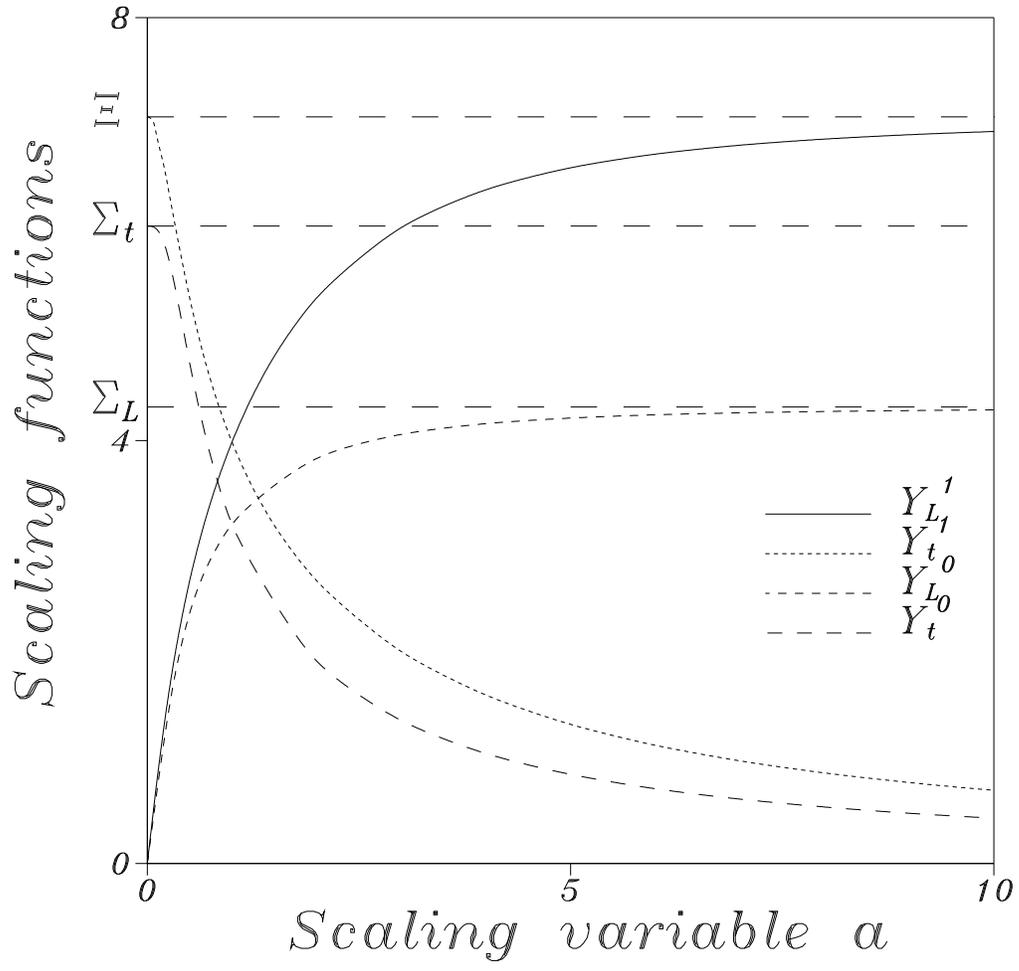}}
\vspace{0.1in}
\caption{The same as in FIG.~\protect\ref{fig2} but for $\lambda=\lambda_{tL}$ and $
a=\frac{tL}{\lambda_{tL}}$. The universal numbers are: $\Xi=7.061132...$, $
\Sigma_{t}=6.028966...$ and $\Sigma_{L}=4.317795...$.}
\label{fig3}
\end{figure}

In Section~\ref{scqp} an analytic continuation of the shift of the critical
quantum parameter for $d=2$ was presented. It is possible to consider the
solutions of Eq.~(\ref{2dcase}) at $\lambda=\lambda_{tL}$ (from Eqs.~(\ref
{lt<1}),~(\ref{tl<1}),~(\ref{tl>1}) and~(\ref{dp=0})) and for different
geometries. In this case the scaling functions $Y_{t}^{1}$, $Y_{L}^{1}$, $%
Y_{t}^{0}$ and $Y_{L}^{0}$ are graphed in FIG.~\ref{fig3}. For $d^{\prime}=1$
again we see that a symmetry between the two limits $a\ll1$ and $a\gg1$ take
place, since the scaling functions $Y_{t}^{1}$ and $Y_{L}^{1}$ are limited
by the universal constant $\Xi$. The asymmetric case $d^{\prime}=0$ has two
different constants $\Sigma_{t}$ and $\Sigma_{L}$, limiting the solutions of 
$Y_{t}^{0}$ and $Y_{L}^{1}$ from above.

The constants $\Xi$, $\Sigma_{t}$ and $\Sigma_{L}$ are obtained from the
asymptotic analysis (with respect to $a$) of Eq.~(\ref{2dcase}) for $%
\lambda=\lambda_{tL}$.

In the limit $a\gg1$ for arbitrary values of $d^{\prime}$ we get (from Eq.~(%
\ref{2da>1})) 
\begin{equation}  \label{a>>1}
B_{0}+\gamma_{E}^{}+\ln\frac{L\phi^{1/2}}{2}= {\sum_{{\bbox l}(2-d^{\prime})}%
}^{\hskip-2mm\prime} K_{0}\left(L\phi^{1/2}|{\bbox l} |\right),
\end{equation}
where the equation of $\lambda_{tL}$ from Eq.~(\ref{gammae}) is used. Eq.~(%
\ref{a>>1}) has the solutions: 
\begin{equation}  \label{XiSigma}
L\chi^{-1/2}=\left\{ 
\begin{array}{ll}
\Xi & \text{for} \ \ \ \ d^{\prime}=1, \\ 
\Sigma_{L} & \text{for} \ \ \ \ d^{\prime}=0,
\end{array}
\right.
\end{equation}
where the universal numbers $\Xi=7.061132...$ and $\Sigma_{L}=4.317795...$
are the solutions of the scaling equation~(\ref{a>>1}) for $d^{\prime}=1$
and $d^{\prime}=0$, respectively.

In the opposite limit $a\ll1$, for $d^{\prime}=1$, we get from Eqs.~(\ref
{tl>1}) and~(\ref{2da<1}) the equation 
\begin{equation}
\gamma_{E}^{}+\ln\frac{\lambda_{tL}\phi^{1/2}}{4\pi t}=2\sum_{m=1}
^{\infty}K_{0}\left(\frac{\lambda_{tL}}{t}\phi^{1/2}m\right),
\end{equation}
which has 
\begin{equation}
\frac{\lambda_{tL}}{t}\chi^{-1/2}=\Xi,
\end{equation}
as a universal solution. For $d^{\prime}=0$ we have 
\begin{equation}  \label{last}
\left(\frac{\lambda_{tL}}{t}\phi^{1/2}-6\right)\exp\left(\frac {\lambda_{tL}%
}{t}\phi^{1/2}\right)- \frac{\lambda_{tL}}{t}\phi^{1/2}-6=0
\end{equation}
obtained from Eqs.~(\ref{dp=0}) and~(\ref{2da<1}), where we have used the
identity~(\ref{ident}).

From Eq.~(\ref{last}) we obtain the universal result 
\begin{equation}  \label{sigmat}
\frac{\lambda_{tL}}{t}\chi^{-1/2}=\Sigma_{t}=6.028966... .
\end{equation}

We finally conclude that if we take $\lambda=\lambda_{c}$ the scaling
functions $Y_{t}^{d^{\prime}}$ and $Y_{L}^{d^{\prime}}$ have a similar
qualitative behavior weakly depending on the geometry (i.e. bloc $%
d^{\prime}=0$ or strip $d^{\prime}=1$) of the system. However, for a given
geometry one distinguishes quite different quantitative behavior of the
scaling functions depending on whether the quantum parameter $\lambda$
is fixed at its critical value, i.e. $\lambda=\lambda_{c}$, or takes
``running'' values $\lambda_{tL}$ obtained from the ``shift equations'' ~(\ref
{lt<1}),~(\ref{tl<1}),~(\ref{tl>1}) or~(\ref{dp=0}).


\subsubsection{Equation of state}\label{subes}
The equation of state of the Hamiltonian~(\ref{model1})
for dimensionalities $1<d<3$ is given by (see Eqs.~(\ref{mag}) and~(\ref
{finite})) 
\begin{eqnarray}
0&=&\delta\lambda- (4\pi)^{-(d+1)/2} \left| \Gamma \left( \frac{1-d}{2}
\right) \right|\left(\frac{h}{{\cal M}} \right)^{(d-1)/2}  \nonumber \\
& &+\frac{\left(\frac{h}{{\cal M}}\right)^{(d-1)/2}}{(2\pi)^{(d+1)/2}} {%
\sum_{m,{\bbox l}(d-d^{\prime})}}^{\hskip-4mm\prime} \ \ \frac{K_{(d-1)/2}
\left[\left(\frac{h}{{\cal M}}\right)^{1/2} \left ((\lambda m/t)^{2} + (L |{%
\bbox l}|)^{2}\right)^{1/2}\right]} {\left[\left(\frac{h}{{\cal M}}%
\right)^{1/2} \left( (\lambda m/t)^{2} + (L |{\bbox l}|)^{2}\right)^{1/2}%
\right] ^{(d-1)/2}}+{\cal M}^{2}.  \label{statel}
\end{eqnarray}
It is straightforward to write this equation in a similar form as in Eq.~(%
\ref{fssl}) or Eq.~(\ref{fsst}), i.e. 
\begin{mathletters}
\label{stateeq}
\begin{equation}
h={\cal M}^{\delta} f_{h}^{L} \left\{\delta\lambda{\cal M}^{-1/\beta}, \frac{%
tL}{\lambda},L^{-1/\nu}{\cal M}^{-1/\beta}\right\},
\end{equation}
or 
\begin{equation}
h={\cal M}^{\delta} f_{h}^{t} \left\{\delta\lambda{\cal M}^{-1/\beta},\frac{%
tL}{\lambda}, \left(\frac{t}{\lambda}\right)^{1/\nu}{\cal M}%
^{-1/\beta}\right\}.
\end{equation}
\end{mathletters}

Eqs.~(\ref{stateeq}) are generalizations of Eqs.~(\ref{bulk16}) in the case
of systems confined to a finite geometry. The appearance of an additional
variable $\frac{tL}{\lambda}$ is a consequence of the fact that the system
under consideration may be regarded as an ``hyperparallelepiped'' (in not
necessary an Euclidean space) of linear size $L$ in $d-d^{\prime}$
directions and of linear size $L_{\tau}$ in one direction with periodic
boundary conditions.

\section{Summary}
Since exact solvability is a rare event in statistical physics, the model
under consideration yields a conspicuous possibility to investigate the
interplay of quantum and classical fluctuations as a function of the
dimensionality and the geometry of the system in an exact manner. Its
relation with the QNL$\sigma$M in the $n\rightarrow\infty$ limit may serve
as an illustration of Stanley's arguments of the relevance of the spherical
approximations in the quantum case. Let us note, however, that the use of
such types of arguments needs an additional more subtle treatment in the finite
size case. For this reason we gave a brief overview of the bulk
low--temperature properties, which are similar to those obtained by
saddle--point calculation for the QNL$\sigma$M (see Section~\ref{bulk}).

The Hamiltonian~(\ref{model1}) can be obtained also in the ``hard--coupling
limit'' from a more realistic model that takes into account the quartic
self--interaction term ${\cal Q}_{\ell}^{4}$, by the ansatz ${\cal Q}%
_{\ell}^{2} \Rightarrow\frac{1}{N}\sum_{\ell}{\cal Q}_{\ell}^{2}$ frequently
used in the theory of structural phase transitions (see Refs.~%
\onlinecite{verbeure92,pisanova93,chamati94,pisanova95} and
Section~\ref{other}).

An analog of the Privman--Fisher hypothesis~\cite{sachdev94,fisher84} for
the FSS form of the free energy in the quantum case was shown to be
consistent with the exact results obtained in Section~\ref{FEN}.
Furthermore, the explicit expression for the universal scaling function
is presented and as a consequence the scaling form of the equation for
the spherical field.

So the Hamiltonian~(\ref{model1}) can be thought of
as a simple but rather general model to test some analytical and numerical
techniques in the theory of magnetic and structural phase transitions.
The discussion of the results obtained, in Section~\ref{bulk}, serves as a
basis for further FSS investigations. Identifying the temperature, which
governs the crossover between the classical and the quantum fluctuations
as an additional temporal dimension one makes possible the use of the
methods of FSS theory in a very effective way.

In Subsection~\ref{scqp} the shift of the critical quantum parameter $%
\lambda $ as a consequence of the quantum and finite--size effects is
obtained. In comparison with the classical case (for details see Ref.~%
\onlinecite{chamati96} and references therein) here the problem is rather
complicated by the presence of two finite characteristic lengths $L$ and $%
L_{\tau}$. We observe a competition between finite size and quantum effects
which reflects the appearance of two regimes: low--temperature ($L\gg
L_{\tau}$) and very low--temperature ($L\ll L_{\tau}$). The behavior of the
shift is analyzed in some actual cases of concrete geometries e.g. strip and
bloc.

In the parameter space (temperature $t$ and quantum parameter $\lambda$),
where quantum zero point fluctuations are relevant, there are three distinct
regions named ``renormalized classical'', ``quantum critical'', and ``quantum
disordered''. The existence of these regions in conjunction with both
regimes: low--temperature and very low--temperature make the model a useful
tool for the exploration of the qualitative behavior of an important class
of systems.

In Subsection~\ref{zerofs} the susceptibility (or the correlation length) is
calculated and the critical behavior of the system in different regimes and
geometries is analyzed. We have studied the model~(\ref{model1}) via $%
\varepsilon$--expansion in order to illustrate the effects of the
dimensionality $d$ on the existence and properties of the ordered phase. An
indicative example is given by Eqs.~(\ref{cy}) and~(\ref{cy1}), while the
former is known (see Ref.~\onlinecite{Zin89}) the last one is quite
different and new. These shows that one must be accurate in taking the limit 
$\varepsilon \rightarrow 0^{+}$ in the context of the FSS theory.

In Subsection~\ref{tdc}, special attention is paid to the two--dimensional
case. The two important cases of strip and bloc geometries are considered.
The universal constant $\Theta$ given by Eq.~(\ref{theta}), which
characterizes the bulk system, is changed to a set of new universal
constants: $\Omega$ (see Eq.~(\ref{Omega})), $\Xi$ and $\Sigma_{L}$ (see
Eq.~(\ref{XiSigma})), and $\Sigma_{t}$ (see Eq.~(\ref{sigmat})). The
appearance of new universal constants reflects the new situation, when there
are two relevant values of the quantum parameter $\lambda$: $%
\lambda=\lambda_{c}$ in the bulk case and $\lambda=\lambda_{tL}$ in the case
of finite geometries. Due to their universality these constants may play an
important role even in studying more complicated models. The
behaviors of the scaling functions at the bulk critical quantum parameter $%
\lambda_{c}$ and the shifted critical quantum parameter $\lambda_{tL}$ are
given in FIGs.~\ref{fig2} and~\ref{fig3}. $L_{\tau}$ is the main
characteristic length and the $\frac{1}{L}$ corrections are exponentially
small in the case of low--temperature regime, and vice versa in the case of
very low--temperature regime.

The equation of state, for the system confined to the general geometry $%
L^{d-d^{\prime}}\!\!\times\!\!\infty^{d^{\prime}}\!\!\times\!\!L_{\tau}$, is
obtained in Subsection~\ref{subes}. This reflects the modifications of the
scaling functions as a consequence of the finite sizes and the temperature.

Finally let us note that this treatment is not restricted to the
Hamiltonian~(\ref{model1}), but it can be applied to a wide class of finite
lattice models (e.g. directly to the anharmonic crystal model see Refs.~%
\onlinecite{verbeure92,pisanova93,chamati94,pisanova95}) and it can also
provide a methodology for seeking different quantum finite--size effects in
such systems.
\acknowledgments
This work is supported by the Bulgarian Science Foundation (projects F608/96
and MM603/96). One of the authors (H.C. would like to thank the
International Atomic Energy Agency and UNESCO for hospitality at the
International Centre for Theoretical Physics, Trieste, where a part of
this paper was performed.
\appendix
\section{Asymptotics of the functions ${\cal K}_{\lowercase{a}} (
\lowercase{\nu|p,y})$}\label{app2}
In this appendix we will sketch a way to find the asymptotic
behavior of the functions ${\cal K}_{a} (\nu|p,y)$ defined in section~\ref
{fing} (see Eq.~(\ref{fssl})). They have the following form 
\begin{mathletters}
\label{b1}
\begin{equation}
{\cal K}_{a} (\nu|p,y) = {\sum_{m,{\bbox l}(p-1)}}^{\hskip-3.5mm\prime} \ \ 
\frac{K_{\nu}\left(2y\sqrt{{\bbox l}^{2}+a^{2}m^{2}}\right)} {\left(y\sqrt{{%
\bbox l}^{2}+a^{2}m^{2}}\right)^{\nu}}, \ \ y>0,
\end{equation}
where 
\begin{equation}  \label{bfq}
{\bbox l}^2 = l_{1}^{2}+l_{2}^{2}+\cdots+l_{p-1}^{2}.
\end{equation}

By the use of the integral representation of the modified Bessel function 
\end{mathletters}
\begin{equation}\label{IR}
K_{\nu}\left(2\sqrt{zt}\right)=K_{-\nu}\left(2\sqrt{zt}\right)= \frac{1}{2}%
\left(\frac{z}{t}\right)^{\nu/2} \int_{0}^{\infty}x^{-\nu-1}e^{-tx-z/x}dx
\end{equation}
and the Jacobi identity for a $p$--dimensional lattice sum 
\begin{equation}  \label{djacobi}
\sum_{m,{\bbox l}(p-1)} e^{-({\bbox l}^{2}+a^{2}m^{2})t}= \frac{1}{a} \left(%
\frac{\pi}{t}\right)^{p/2} \sum_{m,{\bbox l}(p-1)} e^{-\pi^{2} ({\bbox l}%
^{2}+m^{2}/a^{2})/t},
\end{equation}
we may write (\ref{b1}) as 
\begin{eqnarray}
{\cal K}_{a} (\nu|p,y)&=&\frac{\pi^{p/2}}{2a} \Gamma\left(\frac{p}{2}
-\nu\right) y^{-p}  \nonumber \\
& &+\frac{\pi^{2\nu-p/2}}{2a} y^{-2\nu} \int_{0}^{\infty} dx x^{\frac{1}{2}%
p-\nu-1} e^{-x y^{2}/\pi^{2}} \left[{\sum_{m,{\bbox l}(p-1)}}^{\hskip%
-3.5mm\prime} \ \ e^{-x({\bbox l}^{2}+ m^{2}/a^{2})}-a \left(\frac{\pi}{x}%
\right)^{p/2}\right].  \label{A2}
\end{eqnarray}
Let us notice that the two terms in the square brackets in the last equality
cannot be integrated separately, since they diverge. Nevertheless, in order
to encounter this divergence, we can transform further (\ref{A2}) by adding
and subtracting the unity from $\exp(-x y^{2}/\pi^{2})$, which enables us to
write down (after some algebra) the result 
\begin{eqnarray}
{\cal K}_{a} (\nu|p,y)&=&\frac{\pi^{p/2}}{2a} \Gamma\left(\frac{p}{2}
-\nu\right)y^{-p}+\frac{\pi^{2\nu-p/2}}{2a} y^{-2\nu} C_{a}(p|\nu) -\frac{1}{%
2} \Gamma(-\nu)  \nonumber \\
& &+\frac{\pi^{2\nu-p/2}}{2a} \frac{\Gamma\left(\frac{p}{2}-\nu\right)} {%
y^{2\nu}}{\sum_{m,{\bbox l}(p-1)}}^{\hskip-3.5mm\prime} \ \ \left[\left({%
\bbox l}^{2} +\frac{m^{2}}{a^{2}}+\frac{y^{2}}{\pi^{2}} \right)^{\nu-p/2}-%
\left({\bbox l}^{2}+\frac{m^{2}}{a^{2}}\right) ^{\nu-p/2}\right],
\label{k(n,m)}
\end{eqnarray}
where 
\begin{mathletters}
\label{madelung}
\begin{eqnarray}
C_{a}(p|\nu) &=& \lim_{\delta\rightarrow0}\int_{\delta}^{\infty} dx x^{\frac{%
1}{2}p-\nu-1} \left[{\sum_{m,{\bbox l}(p-1)}}^{\hskip-3.5mm\prime} \ \ e^{-x(%
{\bbox l}^{2}+m^{2}/a^{2})}-a \left(\frac{\pi}{x}\right) ^{p/2}\right] , \\
&=&\lim_{\delta\rightarrow0}\left\{{\sum_{m,{\bbox l}(p-1)} }^{\hskip%
-3.5mm\prime}\ \ \frac{\Gamma\left[\frac{p}{2}-\nu,\delta \left({\bbox l}%
^{2}+\frac{m^{2}}{a^{2}}\right)\right]}{\left( {\bbox l}^{2}+\frac{m^{2}}{%
a^{2}}\right)^{\nu-p/2}}\right.  \nonumber \\
& & \ \ \ \ \ \ \ \ \ \ \ \ \ \ \ -
\left.\int_{-\infty}^{\infty}\cdots\int_{-\infty}^{\infty}dmd^{p-1} {\bbox l}%
\frac{\Gamma\left[\frac{p}{2}-\nu,\delta\left({\bbox l}^{2} +\frac{m^{2}}{%
a^{2}}\right)\right]} {\left({\bbox l}^{2}+\frac{m^{2}}{a^{2}}%
\right)^{\nu-p/2}}\right\}
\end{eqnarray}
is the Madelung--type constant and $\Gamma[\alpha,x]$ is the incomplete
gamma function.

We see from Eq.~(\ref{k(n,m)}) that the shift of the critical quantum
parameter is given by the Madelung type constant~(\ref{madelung}) instead of
the sum in Eq.~(\ref{shift}). Indeed it is possible to show that these two
representations are equivalent. This may be done, following Ref.~%
\onlinecite{singh89}, by starting from the Jacobi identity Eq.~(\ref{djacobi}%
), where we multiply the two sides by $\delta^{p/2-\nu-1}$ and integrating
over $\delta$ to obtain the key equation 
\end{mathletters}
\begin{eqnarray}  \label{sp}
C_{a}(p|\nu)&=&{\sum_{m,{\bbox l}(p-1)}}^{\hskip-3.5mm\prime}\ \ \frac{%
\Gamma\left[\frac{p}{2}-\nu,\delta\left({\bbox l}^{2}+\frac{m^{2}} {a^{2}}%
\right)\right]}{\left({\bbox l}^{2}+\frac{m^{2}}{a^{2}}\right) ^{p/2-\nu}}-%
\frac{\delta^{p/2-\nu}}{\frac{p}{2}-\nu}  \nonumber \\
& &\ \ \ +a\pi^{p/2-2\nu}{\sum_{m,{\bbox l}(p-1)}}^{\hskip-3.5mm\prime}\ \ 
\frac{\Gamma\left[\nu,\frac{\pi^{2}\left({\bbox l}^{2}+ a^{2}m^{2} \right)}{%
\delta}\right]}{\left({\bbox l}^{2}+ a^{2}m^{2}\right) ^{\nu}}-a\frac{%
\pi^{p/2}}{\nu \delta^{\nu}}.
\end{eqnarray}

Finally from Eq.~(\ref{sp}) we easily see easily that the integration constant $%
C_{a}(p|\nu)$ may be written in two different forms. In the first case we
take the limit $\delta\rightarrow\infty$ and obtain 
\begin{equation}
C_{a}(p|\nu) = a\pi^{p/2-2\nu}\Gamma(\nu) {\sum_{m,{\bbox l}(p-1)}}^{\hskip%
-3.5mm\prime} \ \ \frac{1}{\left({\bbox l}^{2}+ a^{2}m^{2}\right)^{\nu}}.
\end{equation}
In the other case we take the limit $\delta\rightarrow0$, and then both
first and last terms in the r.h.s of Eq.~(\ref{sp}) yields Eq.~(\ref{madelung}%
).

Using a similar procedure we find, for the functions $\tilde{{\cal K}}%
_{a}(\nu|p,y)$ defined in~(\ref{katt1}), the following expression 
\begin{eqnarray}  \label{lastap}
\tilde{{\cal K}}_{a} (\nu|p,y)&=&\frac{\pi^{p/2}}{2a^{p-1}} \Gamma\left( 
\frac{p}{2}-\nu\right)y^{-p} +\frac{\pi^{2\nu-p/2}}{2a^{p-1}} y^{-2\nu} 
\tilde{C}_{a}(p|\nu) -\frac{1}{2} \Gamma(-\nu)  \nonumber \\
& &+\frac{\pi^{2\nu-p/2}}{2a^{p-1}} \frac{\Gamma\left(\frac{p}{2}-\nu \right)%
}{y^{2\nu}}{\sum_{m,{\bbox l}(p-1)}}^{\hskip-3.5mm\prime}\ \ \left[\left(%
\frac{{\bbox l}^{2}}{a^{2}} +m^{2}+\frac{y^{2}} {\pi^{2}}\right)^{\nu-p/2}-%
\left(\frac{{\bbox l}^{2}}{a^{2}} +m^{2}\right)^{\nu-p/2}\right].
\end{eqnarray}
Here the Madelung type constant is given by 
\begin{mathletters}
\label{madelungt}
\begin{eqnarray}
\tilde{C}_{a}(p|\nu)&=&\lim_{\delta\rightarrow0}\int_{\delta}^{\infty} dx x^{%
\frac{1}{2}p-\nu-1}\left[{\sum_{m,{\bbox l}(p-1)}}^{\hskip-3.5mm\prime}\ \
e^{-x({\bbox l}^{2}/a^{2}+m^{2})}- a^{p-1} \left(\frac{\pi}{x}%
\right)^{p/2}\right] , \\
&=&\lim_{\delta\rightarrow0}\left\{{\sum_{m,{\bbox l}(p-1)} }^{\hskip%
-3.5mm\prime}\ \ \frac{\Gamma\left[\frac{p}{2}-\nu,\delta\left( \frac{{\bbox %
l}^{2}}{a^{2}}+m^{2}\right)\right]}{\left(\frac{{\bbox l} ^{2}}{a^{2}}%
+m^{2}\right)^{\nu-p/2}} \right.  \nonumber \\
& & \ \ \ \ \ \ \ \ \ \ \ \ \ \ \ -
\left.\int_{-\infty}^{\infty}\cdots\int_{-\infty}^{\infty}dmd^{p-1} {\bbox l}%
\frac{\Gamma\left[\frac{p}{2}-\nu,\delta\left(\frac{{\bbox l} ^{2}}{a^{2}}%
+m^{2}\right)\right]}{\left(\frac{{\bbox l}^{2}}{a^{2}} +m^{2}\right)^{%
\nu-p/2}}\right\} \\
&=&a^{p-1}\pi^{p/2-2\nu}\Gamma(\nu){\sum_{m,{\bbox l}(p-1)} }^{\hskip%
-3.5mm\prime} \ \ \frac{1}{\left(a^{2}{\bbox l}^{2}+ m^{2}\right)^{\nu}}.
\end{eqnarray}

Eqs.~(\ref{k(n,m)}) and~(\ref{lastap}) are slight generalizations (for the
anisotropic case $a\neq1$) of the result obtained in Ref.~%
\onlinecite{singh89} from one side, and get in touch with the Watson type
sums proposed earlier in Ref.~\onlinecite{brankov88} from the other (see
also Ref.~\onlinecite{chamati96}).

If we put in Eqs.~(\ref{k(n,m)}) or~(\ref{lastap}) $d=2$, $d^{\prime}=0$ and 
$a=1$ we obtain the identity 
\end{mathletters}
\begin{eqnarray}  \label{ident}
{\sum_{l_{1},l_{2}}}^{\prime}\frac{\exp\left(-y \sqrt{l_{1}^{2}+l_{2}^{2}}%
\right)}{\sqrt{l_{1}^{2}+l_{2}^{2}}}&=& \frac{2\pi}{y}+4\zeta\left(\frac{1}{2%
}\right) \beta\left(\frac{1}{2}\right)+y  \nonumber \\
& &+2\pi{\sum_{l_{1},l_{2}} }^{\prime}\left\{\frac{1}{\sqrt{%
y+4\pi^{2}(l_{1}^{2}+ l_{2}^{2})}}-\frac{1}{2\pi\sqrt{l_{1}^{2}+l_{2}^{2}}}%
\right\}.
\end{eqnarray}

\section{Shift of the critical quantum parameter for some particular
geometries}\label{app3}

Our task in this appendix is to explain how the shift of the
critical quantum parameter in the special cases $d=2$ (for the
low--temperature regime) Eq.~(\ref{gammae}) and $d^{\prime}=1$ (for the very
low--temperature regime) Eq.~(\ref{d'=1}) are obtained.

\subsection{Low--temperature regime, $d=2$}

The expression for the shift is given by Eq.~(\ref{shift1}) for an arbitrary
dimensionality of the system. To calculate the shift for $d=2$ we proceed as
follows: we first put $d=2+\varepsilon$ and expand the obtained expressions
around $\varepsilon=0$. So, for the functions in the first term of Eq.~(\ref
{shift1}) we obtain 
$$
\pi^{-(3+\varepsilon)/2}=\pi^{-3/2}\left(1-\frac{\varepsilon}{2}
\ln\pi\right)+{\cal O}(\varepsilon^{2}), 
$$
$$
\left(\frac{t}{\lambda_{tL}}\right)^{1+\varepsilon}= \frac{t}{\lambda_{tL}}%
\left(1+\varepsilon\ln\frac{t}{\lambda_{tL}} \right)+{\cal O}%
(\varepsilon^{2}), 
$$
$$
\Gamma\left(\frac{1+\varepsilon}{2}\right)=\sqrt{\pi}\left[1- \frac{%
\varepsilon}{2}(2\ln2+\gamma_{E}^{})\right]+{\cal O} (\varepsilon^{2}), 
$$
$$
\zeta(1+\varepsilon)=\frac{1}{\varepsilon}+\gamma_{E}^{}+ {\cal O}%
(\varepsilon), 
$$
where $\gamma_{E}^{}$ is Euler's constant. For the second term we obtain 
$$
L^{-\varepsilon}=1-\varepsilon\ln L+{\cal O}(\varepsilon^{2}), 
$$
$$
\Gamma\left(\frac{\varepsilon}{2}\right)=\frac{2}{\varepsilon} \left[1-\frac{%
\varepsilon}{2}\gamma_{E}^{}\right]+{\cal O} (\varepsilon), 
$$
and 
\begin{equation}  \label{AB1}
\lim_{d\rightarrow 2+\varepsilon} {\sum_{{\bbox l}(d-d^{\prime})} }^{\hskip%
-2mm\prime}|{\bbox l}|^{2-d}\equiv{\cal Z}\left| 
\begin{array}{c}
0 \\ 
0
\end{array}
\right| ({\bbox l},\varepsilon)=-1+B_{0}\varepsilon+{\cal O}%
(\varepsilon^{2}).
\end{equation}
Finally after the substitution of the above expansions in Eq.~(\ref{shift1})
and then taking the limit $\varepsilon\rightarrow0$ we obtain
Eq.~(\ref{gammae}).

Let us now consider the particular cases $d^{\prime}=1$ and $d^{\prime}=0$

{\it The case $d^{\prime}=1$}: Here the ($d-d^{\prime}$)--dimensional sum in
Eq.~(\ref{AB1}) reduces to~\cite{ryzhik80} 
\begin{equation}  \label{exzeta}
{\sum_{l}}^{\prime} l^{-\varepsilon}=2\zeta(\varepsilon) = -1 -
\varepsilon\ln(2\pi)+{\cal O}(\varepsilon^{2}),
\end{equation}
hence the constant $B_{0}=-\ln2\pi$ and the result~(\ref{lt<1}) emerges.

{\it The case $d^{\prime}=0$}: Now the sum in Eq.~(\ref{AB1}) is
two--dimensional. Then we obtain the following result due to
Hardy~\cite{glasser80}
\begin{equation}  \label{hardi}
{\sum_{l_{1},l_{2}}}^{\prime} \left(l_{1}^{2}+
l_{2}^{2}\right)^{-\varepsilon/2}
=4\zeta\left(\varepsilon/2\right)\beta\left(\varepsilon/2\right),
\end{equation}
where 
$$
\beta(s)=\sum_{l=1}^{\infty}\frac{(-1)^{l}}{(2l+1)^{s}}, 
$$

To obtain the expansion of the functions in Eq.~(\ref{hardi}) we use~(\ref
{exzeta}) and 
\begin{equation}
\beta(\varepsilon)= 4^{-\varepsilon}\left\{\zeta\left(\varepsilon,\frac{1}{4}%
\right)-\zeta \left(\varepsilon,\frac{3}{4}\right)\right\} =\frac{1}{2}+%
\frac{\varepsilon}{2}\ln\frac{\left[\Gamma \left(\frac{1}{4}%
\right)\right]^{2}} {2\pi\sqrt{2}}+{\cal O}(\varepsilon^{2}),
\end{equation}
obtained with the aid of the expansion~\cite{ryzhik80} 
$$
\zeta(\varepsilon,a)=\frac{1}{2}-a+\varepsilon\left[\ln\Gamma(a)- \frac{1}{2}%
\ln2\pi\right]+{\cal O}(\varepsilon^{2}), 
$$
for the incomplete zeta function $\zeta(s,a)$. Finally we get 
\begin{equation}
{\sum_{l_{1},l_{2}}}^{\prime} \left(l_{1}^{2}+l_{2}^{2}
\right)^{-\varepsilon/2} =-1-\varepsilon\ln\frac{\left[\Gamma\left(\frac{1}{4%
}\right) \right]^{2}}{2\sqrt{\pi}}+{\cal O}(\varepsilon^{2}),
\end{equation}
which makes possible the calculation of the shift, Eq.~~(\ref{tl<1}), of the
critical quantum parameter for the fully finite geometry.

\subsection{Very low--temperature regime, $d^{\prime}=1$}

In this limit the shift is given by Eq.~(\ref{shift2}). Here we will set $%
d^{\prime}=1+\varepsilon$ and expand around $\varepsilon=0$. In the first
term of Eq.~(\ref{shift2}) we have a $(d-d^{\prime})$--dimensional sum,
which can be replaced by the corresponding Epstein zeta function. Then we
can use the following asymptotic form~\cite{glasser80} 
\begin{equation}
{\sum_{{\bbox l}(d-d^{\prime})}}^{\hskip-2mm\prime} |{\bbox l}|^{1-d}\equiv 
{\cal Z}\left| 
\begin{array}{c}
0 \\ 
0
\end{array}
\right| ({\bbox l}^{2},d-1)=\frac{2\pi^{(d-1)/2}}{\Gamma\left(\frac{d-1}{2}
\right)\varepsilon}+C_{0}+{\cal O}(\varepsilon), \ \ \ \ \ \ \
d^{\prime}=1+\varepsilon,
\end{equation}
where the expressions for the constants $C_{0}$ are quite complicated except
for some special cases (see Refs.~\onlinecite{singh89}, e.g. for $d=2$, $%
d^{\prime}=1-\varepsilon$ we have $C_{0}=\gamma_{E}^{}-\ln(4\pi)$ (c.f. Ref.~%
\onlinecite{Zin89}, Eq. (30.104)).

Further for the function in the second term of Eq.~(\ref{shift2}) we use the
expansions given in the preceding subsection. After substituting the
obtained expansions in the basic expression~(Eq.(\ref{shift2})) for the
shift, we obtain Eq.~(\ref{d'=1}).

\begin{references}
\bibitem{dyn} In this approximation the dynamical critical exponent $z$ is
equal to the unity i.e. there is no anisotropy between the space and time
dimensions.
\bibitem{index}  The subscripts $T$ and $\lambda $ mean that the critical
exponents are related with the change of $t$ or $\lambda $, respectively.
\end{references}
\end{document}